\documentclass[%
 reprint,
 amsmath,amssymb,
 aps,
]{revtex4-2}

\usepackage[usenames,dvipsnames]{xcolor} 
\usepackage{graphicx}
\usepackage{mathtools}
\usepackage{dcolumn}
\usepackage{bm}
\usepackage[colorlinks=true,citecolor=blue,linkcolor=blue]{hyperref}
\usepackage[caption=false]{subfig}
\usepackage{bbm}
\usepackage{braket}
\usepackage{comment}
\usepackage{hyperref}
\usepackage{upgreek}
\usepackage{esint}
\usepackage[normalem]{ ulem } 
\usepackage{todonotes}

\begin{document}

\preprint{APS/123-QED}

\title{Fluctuations in heat current and scaling dimension}

\author{Hiromi Ebisu}
\affiliation{\mbox{Department of Condensed Matter Physics, Weizmann Institute of Science, Rehovot 76100, Israel}}
\author{Noam Schiller}
\affiliation{\mbox{Department of Condensed Matter Physics, Weizmann Institute of Science, Rehovot 76100, Israel}}
\author{Yuval Oreg}
\affiliation{\mbox{Department of Condensed Matter Physics, Weizmann Institute of Science, Rehovot 76100, Israel}}

\begin{abstract}
In this work, we theoretically study the heat flow between two
$1+1$d chiral gapless systems
connected by a point contact. With a small temperature gradient between the two, we find that the ratio between fluctuations of the heat current and the heat current itself is proportional to the scaling dimension -- a universal number that characterizes the distribution of the particles tunneling through the point contact. We adopt two different approaches, scattering theory and conformal field theory, to calculate this ratio and see that their results agree. Our findings are useful for probing not only fractional charge excitations in fractional quantum Hall states but also neutral ones in non-Abelian phases. 
\end{abstract}

\maketitle

\textit{Introduction.}--
%
The flow of charge in an electric circuit is not continuous due to the discrete nature of the charge carriers. Similarly, we expect that the energy or the heat flow in the system will be comprised of ``lumps" of energy associated with each carrier.
In 1918, Schottky investigate the non-continuous character by measuring electric current fluctuations in a vacuum tube \cite{schottky_uber_1918}. In these tubes, the current flows by Poisson processes of independent and rare electron emission events from the tube filament. The properties of the Poisson distribution indicate that the ratio between the variance of the electric current and the average current is equal to the electric charge emitted by a single event – the electron charge. 

There is a clear physical distinction between electric current and heat current: while in each random emission event the charge is fixed, the energy carried by the emitted carrier is not. 
Furthermore, the probability of emission itself may depend on the energy. 
The ratio between fluctuations in the electric current and the average electric current have long been used to infer the charge of carriers~\cite{de-picciotto_direct_1997,saminadayar_observation_1997,Griffiths2000,Dolev2008}. Analogously, it stands to reason that the ratio between the fluctuations in the heat current and the average heat current can be utilized to infer properties of the energy distribution of the emission events. In this manuscript, we show that this is indeed the case. In particular, the quantum number that can be extracted from these measurements is the scaling dimension of the emitted quasiparticles, $h$.

We study the heat flow between two 
$1+1$d chiral gapless systems 
connected by a point contact. A canonical example of such systems are the edges of two-dimensional systems subjected to a strong magnetic field in the fractional quantum Hall (FQH) regime. Quasiparticles which tunnel between the edges carry charge, as well as energy. 
Much experimental progress has been made recently in heat current measurement of FQH states, including non-Abelian phases~\cite{jezouin2013quantum,banerjee_observed_2017,banerjee_observation_2018,Matsuda2018}.
Theories of FQH states predict the charges of the quasiparticles $e^*$ as well as the scaling dimension $h$ of the operator creating a quasiparticle~\cite{witten1989quantum,ELITZUR1989108,wen_quantum_2004}. In certain simple cases, for example at $\nu=1/3$, there is a simple relation between $h$ and, $\theta$, the exchange statistics phase of the quasiparticles, via $e^{i\theta}=e^{2\pi ih}$. The scaling dimension also determines the exponent of the power law of the effective tunneling amplitude as a function of energy~\cite{kane_transport_1992,kane_transmission_1992}. Previous attempts to measure the scaling dimension focused on measuring these power laws through the charge current~\cite{Radu_experiment,Baer_experiment,lin_measurements_2012}.

We calculate the average heat current which tunnels through a quantum point contact (QPC), $I_E$, as well as the fluctuations of the heat current, $S_E$. When the temperature difference between the two edges, $\Delta T$, is much smaller than the temperature of the cold edge, $T$, we find that the ratio between the two is
\begin{equation}
    \label{eq:Close_Temp}
    \mathcal{F}_E=\frac{S_E}{2I_E}=(4h+1) k_B T,
\end{equation}
where $k_B$ is the Boltzmann constant. This universal result is valid also for charge-neutral particles and  demonstrates the importance of the scaling dimension in governing the energy distribution of excitations along the edge. We emphasize the difference between this regime and the shot-noise regime typically used to extract quasiparticle charge~\cite{de-picciotto_direct_1997,saminadayar_observation_1997,Griffiths2000,Dolev2008}, which requires a \textit{large} bias voltage $V \gg T$ between the two edges.

In the rest of the paper we will prove equation Eq.~\eqref{eq:Close_Temp} and show that it holds for generic interacting edge modes described by conformal field theory (CFT)~\cite{DiFrancesco1997}. 
We extend these results beyond the small-$\Delta T$ regime described in Eq.~\eqref{eq:Close_Temp}, obtaining a closed integral expression for the heat current and heat current fluctuations for general values of $T_L$ and $T_R$. We focus on another limit of interest, tunneling from a hot edge $T_L=T$ to a cold edge $T_R \rightarrow 0$, in Eq.~\eqref{integral_ex_compact}. 

A schematic demonstration of how quantum dots may be used to probe the heat current fluctuations is given in App.~\ref{App:schematic_quantum_dot} for non-interacting fermionic states. Generalization to other cases requires more delicate treatment in the spirit of, for example, Ref.~\cite{furusaki_resonant_1998}, which we leave for future works.

\begin{figure}
    \centering
    \includegraphics[width=\columnwidth]{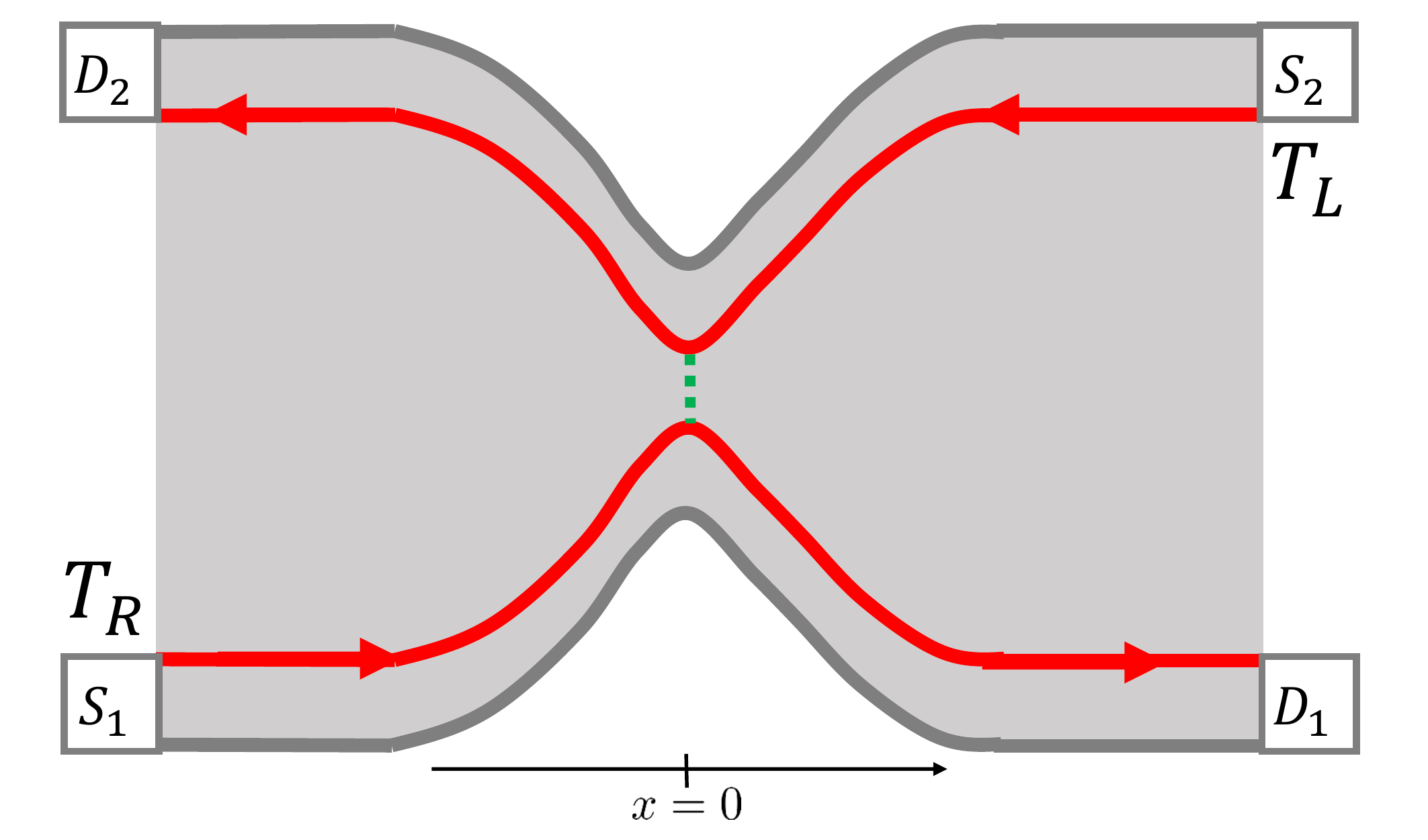}
    \caption{
    A configuration of counter-propagating edge modes (red arrows) of an integer quantum Hall state. The quantum point contact (QPC), where quasiparticle tunneling occurs (green dashed line), is located at $x=0$. The source and drain of the right-/left-moving edge mode are respectively represented by $S_{1/2}$ and~$D_{1/2}$. Depending on the context, the edge modes can be replaced with the ones of a fractional quantum Hall state with filling fraction $\nu$ or with the ones of topological ordered phases described by a conformal field theory. }
    \label{fig01}
\end{figure}

 \textit{Scattering theory}-- We begin with calculations for non-interacting fermions and bosons, using the standard tool of scattering theory. Focusing on the geometry of Fig.~\ref{fig01}, the heat current and heat current fluctuations measured at drain $1$ are given by the standard formulas ~\cite{blanter_shot_2000,krive_heat_2001} 
\begin{align}
    \label{eq:Scattering_QPC_I}
    I_{E} & =\frac{1}{h}\int dER\left(E\right)\left[f^{\pm}_{R}-f^{\pm}_{L}\right]E, \\
    \tilde{S}_{E} & =\frac{2}{h} \int dE E^{2} \left\{ \left( 2-R\left(E\right) \right) f^{\pm}_{R} \left[1\mp f^{\pm}_{R}\right] \right.
            \label{eq:Scattering_QPC_S}\\
     & \!\!\!\!\!\!\!\! \left.+R\left(E\right)f^{\pm}_{L} \left[1 \mp f^{\pm}_{L}\right]\pm R\left(E\right)\left(1-R\left(E\right)\right)\left(f^{\pm}_{R}-f^{\pm}_{L}\right)^{2}\right\} .\nonumber 
\end{align}
Here, $f^{\pm}_{R/L} \equiv \left[ \exp \left(E/T_{R/L}\right) \pm 1 \right] ^{-1}$ is the distribution function for the right-/left-moving particles, with the $+$ sign representing the Fermi-Dirac distribution for fermions, and the $-$ sign representing the Bose-Einstein distribution for bosons. The coefficient $R \left(E\right)$ is the probability for an incident particle of energy $E$ to tunnel across the QPC. We integrate Eqs.~(\ref{eq:Scattering_QPC_I}) and (\ref{eq:Scattering_QPC_S}) for a general scatterer obeying $R \left( E \right) = R_0 \left(E/E_c\right)^\alpha $, where $E_c \gg T_i$ is a high-energy cutoff. As will be shown in the following sections, for a standard QPC, we expect $\alpha = 4 h -2$, where $h$ is the scaling dimension of the tunneling quasiparticle. This corresponds to $\alpha = 0$ for fermions and $\alpha = 2$ for bosons.

We wish to use these quantities to probe the quasiparticles which tunnel through the scatterer. As such, we focus only on the excess heat current fluctuations, \textit{i.e.,} the noise contribution that is obtained from a non-zero $R(E)$, $S_E \equiv \tilde{S}_E-\tilde{S}_E \big|_{R(E)=0}$. We then define the ``heat Fano factor" as the ratio between the excess heat current fluctuations and the tunneling heat current, $\mathcal{F}_E \equiv S_E/2 I_E$.

Calculating these integrals, we obtain the heat Fano factor for two limits of interest. For nearby temperatures, $T_R \equiv T,  T_L =T+\Delta T$, we obtain, for both fermions and bosons, 
\begin{equation}
    \label{eq:Scattering_Close}
    \mathcal{F}_E =  \left( \alpha + 3 \right) k_B T.
\end{equation}
The limit of $T_L=T$ and $T_R=0$ and further details are relegated to  App.~\ref{App:scattering_theory}.
\par
 \textit{CFT approach}-- 
 Now we turn to \textit{interacting} cases by considering the edge modes described by a CFT with central charge $c$. This approach enables us to study not only simple Abelian phases, such as FQH states at filling fraction $\nu=1/m$, with $m$ being an odd integer, but also
 non-Abelian topological phases. Two prominent such phases, the Moore-Read state~\cite{MOORE1991362} and the
 $SU(2)_k$ topological order phases (chiral spin liquid phases)~\cite{Deser1982,KLstate1987}, host neutral quasiparticle excitations, emphasizing the benefits of considering the heat Fano factor $\mathcal{F}_E$ vs. the more standard charge Fano factor. 
 We remark that, in the case of Abelian FQH states, one can obtain the same 
 results by means of the standard bosonization formalism instead of adapting the generic CFT approach. This is illustrated in detail in App.~\ref{App:luttinger_liquid}.

 For simplicity, in the following calculations, we will set $\hbar=k_B=1$.
We envisage the same  geometry as the scattering theory portrayed in Fig.~\ref{fig01}, 
  where there is a pair of counter-propagating edge modes of a CFT, connected by a single QPC. We denote the temperature of the right-/left-moving edge by $T_{R/L}$, and introduce the QPC at the coordinate $x=0$.
Notice that we do not impose a potential gradient between the two edges.
  
The Hamiltonian of this system is given by
  \begin{eqnarray}
H&=&H_0+H_T,\nonumber\\
  H_{0}&=&\frac{v}{2\pi}\int _{-\infty}^{+\infty}dx[\mathcal{T}+\overline{\mathcal{T}}]\nonumber\\
      H_{T}&=& \Gamma_0 O_R(0)O_L(0).\label{076}   
\end{eqnarray}
Here, $H_0$ describes kinetic terms consisting of the stress-energy tensor $\mathcal{T}(z)(\overline{\mathcal{T}}(\overline{z}))$ in the right (left) moving 
sector with $z=i(vt-x)(\overline{z}=i(vt+x))$  and $v$ is the velocity of the edge mode~\cite{DiFrancesco1997}. The term $H_T$ represents tunneling at the QPC, and the tunneling entity is described by the operator $O_{R/L}(z)$ with a tunneling amplitude $\Gamma_0$. This operator is a primary field of the CFT, with a scaling dimension of $h_{O_R}=h_{O_L}\equiv h_{O}$~\cite{TechnicalCaveat}.

We evaluate the heat current at the drain $D_1$, which we place at coordinate $x=d$, at time $t$.  
The unperturbed heat current is defined by $I_E^{(0)}(t) = \frac{v^2}{2\pi} \mathcal{T}(d,t)$~\cite{CAPPELLI2002568}. To find the correction to the heat current induced by the tunneling at the QPC, we resort to linear response theory, assuming the tunneling $H_T$ in Eq.~\eqref{076} is perturbatively small and the perturbation $H_T$ is turned on at time~$t=-\infty$. One obtains the perturbative expansion of the heat current operator up to the second order of $\Gamma_0$:
  \begin{equation}
I_E(t)=I^{(0)}_E(t)+I^{(1)}_E(t)+I^{(2)}_E(t)+
\mathcal{O}(\Gamma_0^3), \label{011th2}
\end{equation}
where 
\begin{align*}
   I_E^{(1)}(t)=  i & \int^t_{-\infty}dt^{\prime}[H_T(t^{\prime}),\frac{v^2}{2\pi}\mathcal{T}(d,t)], \\
  I_E^{(2)}(t)=  i^2\int^t_{-\infty}dt^{\prime} & \int^{t^{\prime}}_{-\infty}dt^{\prime\prime}\bigl[H_T(t^{\prime\prime}),[H_T(t^{\prime}),\frac{v^2}{2\pi}\mathcal{T}(d,t)]\bigr].
\end{align*}
See App.~\ref{App:generic_cft} for more details.

The unperturbed heat current $I^{(0)}_E(t)$ is related to the heat conductance via $\kappa=\frac{\partial \braket{I^{(0)}_E(t)}}{\partial T_R}=\frac{\pi c}{6}T_R$, from which we can extract the central charge $c$ of the edge mode~\cite{pendry1983quantum,kane_quantized_1997,CAPPELLI2002568,banerjee_observed_2017,banerjee_observation_2018}. Since we are interested in excess heat fluctuations driven by the quasiparticle tunneling, throughout this work, we concentrate on the perturbative corrections to the heat current and noise, \textit{i.e.,} the second and third term in Eq.~\eqref{011th2}.
Taking the expectation value of these terms gives
\begin{eqnarray}
  \braket{I^{(1)}_E(t)}&=&
  0,\nonumber\\
    \braket{I^{(2)}_E(t)}&=&-i\Gamma_0^2\int_{-\infty}^{+\infty}d\tau G_R(\tau)\partial_{\tau}G_L(\tau),
   \label{010th}
\end{eqnarray}
  where we have introduced the correlator of the primary field in right-/left-moving sector at temperature $T_{R/L}$ as 
  $G_{R/L}(\tau)=\braket{O_{R/L}(\tau)O_{R/L}(0)}$. The scaling dimension enters as the power law of the correlator. At zero temperature, this can be seen as $G_{R/L}(\tau)\sim\tau^{-2h_{O_{R/L}}}$.

\par
When there is no temperature gradient between the edges, $\braket{I_E^{(2)}(t)}$ vanishes~\footnote{When the temperatures of both edges are equal, \textit{i.e.,} $T_R=T_L$, the correlator $G_R(\tau)$ is equivalent to $G_L(\tau)$. Hence, $ \braket{I^{(2)}_E(t)}$ vanishes, as the integrand is proportional to total derivative of $G_L^2(\tau)$. This is physically plausible since the geometry we consider becomes symmetric when $T_R=T_L$ and the amount of heat transferred from right to left edge through the QPC is precisely compensated by the one from left to right, leading to the zero net heat transfer.}.
In the presence of a small  
temperature gradient between the two edge modes, \textit{i.e.,} $T_R=T$, $T_L=T+\Delta T$ ($|\Delta T/T|\ll 1$), the heat current~\eqref{010th} behaves as $\braket{I^{(2)}_E(t)}\sim T^{4h_O-1}\Delta T$.
\par
  We proceed to calculate the heat current fluctuations. We define these as 
  \begin{equation*}
    \tilde{S}_E(\omega)=\int_{-\infty}^{+\infty}dt_{12}e^{i\omega t_{12}}\braket{\{\Delta I_E(t_1),I_E(t_2)\}},
\end{equation*}
with $\Delta I_E(t)=I_E(t)-\braket{I_E(t)}$ and $\{\cdot,\cdot\}$ representing an anti-commutator.
Expanding the heat current up to second order in $\Gamma_0$,
as demonstrated in Eq.~\eqref{011th2},  yields~\footnote{One can prove $S_E^{(10)}(\omega)=S_E^{(01)}(\omega)=0$ by an  analogous argument to the one which gives $\braket{I^{(1)}_E(t)}=0$ in Eq.~\eqref{010th}. Hence, these terms are omitted.}
\begin{equation*}
    \tilde{S}(\omega)=S_E^{(00)}(\omega)+S_E^{(11)}(\omega)+S_E^{(02)}(\omega)+S_E^{(20)}(\omega)+\mathcal{O}(\Gamma_0^3).
\end{equation*}
Here, we have defined $(i,j=0,1,2)$
\begin{equation*}
    S_E^{(ij)}(\omega)=\int_{-\infty}^{+\infty}dt_{12}e^{i\omega t_{12}}\braket{\{\Delta I^{(i)}_E(t_1),I^{(j)}_E(t_2)\}}
\end{equation*}

The term $S_E^{(00)}(\omega)$ gives the unperturbed heat current fluctuations, and is an equilibrium property. Known also as the so-called Johnson-Nyquist (JN) noise of the heat current~\cite{Johnson,Nyquist}, it is related to the heat conductance $\kappa$ in the DC limit by $S_E^{(00)}(0)=2\kappa T_R^2$, which is addressed for non-interacting cases in Ref.~\cite{krive_heat_2001}. This relation is in line with the one between the charge JN noise $S_c$ and the charge conductance $G$ via $S_c=2GT$ in the thermal noise limit.\par
We are interested only in the perturbative contributions to the heat current fluctuations. We hence focus on the excess heat current fluctuations, defined as $S_E(\omega)=\tilde{S}_E(\omega)-S_E^{(00)}(\omega)$. The excess heat current fluctuations are comprised of three terms. For $S_E^{(11)}(\omega)$, which represents the auto-correlations of the heat current which tunnels at the QPC, 
it is straightforward to derive the form
\begin{equation}
     S_E^{(11)}(\omega)=-2i\Gamma_0^2\int^{+\infty}_{-\infty}d\tau \cos \left(\omega\tau \right) G_R(\tau)\partial_{\tau}^2G_L(\tau).\nonumber
\end{equation}

To evaluate the  ``cross terms"  $S_E^{(02)}(\omega)+S_E^{(20)}(\omega)$, corresponding to the correlation between the unperturbed and excess heat current, one has to calculate the correlator involving the stress-energy tensor and the primary fields. Such a task can be accomplished
by exploiting the conformal Ward identity~\cite{DiFrancesco1997},
as outlined in App.~\ref{App:generic_cft}. 
Overall, the excess heat current fluctuations in the DC limit $\omega\to0$ is given by
\begin{align}
    \label{cshot noise main}
   S_E(0) & =
    4T_R(2h_O-1)\braket{I_E^{(2)}} \\
    & +S_E^{(11)}(0)+2iT_R\frac{\partial S_E^{(11)}(\omega)}{\partial \omega}\Bigg|_{\omega\to0}. \nonumber
\end{align}
  
  We are now are in a good place to study the heat Fano factor, defined by
  \begin{equation*}
      \mathcal{F}_E=\frac{S_E(0)}{2\braket{I^{(2)}_E(t)}}.
  \end{equation*}
The first term of Eq.~\eqref{cshot noise main} includes $\braket{I^{(2)}_E(t)}$, therefore, the heat Fano factor has the term proportional to the scaling dimension which is a universal number. What remains to do is to calculate the ratio between the last two terms of Eq.~\eqref{cshot noise main} and $\braket{I^{(2)}_E(t)}$.\par
It is challenging to evaluate this ratio analytically for generic values of $T_{R/L}$. Instead of doing this, we focus on the case of a small temperature gradient, $T_R=T$, $T_L=T+\Delta T$ ($|\Delta T/T|\ll 1$), and expand the last two terms in Eq.~\eqref{cshot noise main} up to first order of $\Delta T/T$.
Relegating the details to App.~\ref{App:generic_cft}, and retrieving the Boltzmann constant $k_B$, the heat Fano factor becomes $\mathcal{F}_E=(4h_O+1)k_BT$, which completes the proof of
Eq.~\eqref{eq:Close_Temp}.\par
  We extend our result to an additional regime of interest with a large temperature gradient,
 by setting $T_R=0$ and $T_L=T$. 
In this case, the heat Fano factor is given by
\begin{equation}
   \mathcal{F}_E=(2h_O+1)(\pi k_BT)
   \frac{J \left(2 h_O,2 \right)}{J \left(2 h_O,1 \right)},
    \label{integral_ex_compact}
\end{equation}
where we define the integral (see App.~\ref{App:generic_cft}) 
\begin{equation*}
    J \left(a,b\right) =  \int^{+\infty}_{-\infty}dz
    \left[ \cosh(z) \right]^{-a} 
    \left[i \left( z-i\frac{\pi}{2} \right) \right]^{- a - b}.
\end{equation*}
When $h_O$ takes integer or half-integer values, the integrals of Eq.~\eqref{integral_ex_compact} can be explicitly calculated via contour integration. This includes two canonical cases: when~$h_O=1/2$, corresponding to fermion tunneling, the heat Fano factor is given by $\mathcal{F}_E=3k_BT\frac{\zeta(3)}{\zeta(2)}$; when $h_O=1$, which corresponds to density-density interactions, we have $\mathcal{F}_E=4k_BT\frac{\zeta(5)}{\zeta(4)}$.  
Here, $\zeta(s)$ is the Zeta function. 
Interestingly, this result coincides with the one obtained from scattering theory. The more general case is relegated to App~\ref{App:generic_cft}. \par 

\textit{Discussion}--
We identify several advantages to focusing on heat vs. alternate approaches that have been proposed to extract the scaling dimension from charge fluctuations \cite{shtanko_nonequilibrium_2014,snizhko_scaling_2015,rech_negative_2020,schiller_extracting_2021}. 
First, a crucial feature of the heat current measurement is that it may probe not only charged excitations but also \textit{neutral} ones, as both of them carry heat. 
Heat current measurement thus opens a new possibility to provide us with smoking gun evidence for neutral topological order phases~\cite{Deser1982,KLstate1987,KITAEV20062}. Indeed, experiments probing the quantized heat conductance $\kappa=c\frac{\pi k_B}{6 \hbar}T$, with $c$ being the central charge of the FQH edge mode, have been successfully realized for both integer~\cite{jezouin2013quantum,banerjee_observed_2017} and half-integer~\cite{banerjee_observation_2018,Matsuda2018} central charges, the latter hosted by non-Abelian phases~\cite{MOORE1991362}.\par
Second, charge current based measurements often require bias voltages that are \textit{larger} than the base temperature. This, combined with the requirement that voltages not exceed the bulk gap, limits the available temperature range for measurements. Large bias voltages may furthermore lead to non-universal effects, such as changing the confining potential at the edge and the electrostatic properties of the QPC, or enabling longitudinal conductance through the bulk. Conversely, our method requires only small temperature gradients, such that effects that may mask the power law behavior of the tunneling are minimized. Finally, probing power laws directly requires bias voltages which span several decades, an issue which we outright avoid.

\textit{Summary}--
Recent decades have seen considerable interest in noise as a means for extracting physical insight, as opposed to an unfortunate byproduct of an imperfect measuring apparatus. In the condensed matter community, the ratio between the electric current fluctuations and the average electric current has been used to extract the quantized charge of carriers~\cite{de-picciotto_direct_1997,saminadayar_observation_1997,Griffiths2000,Dolev2008}. In this paper, we examine the ratio between the heat current fluctuations and the average heat current, which we dub the heat Fano factor. We refer the reader to further works discussing heat current fluctuations in theoretical \cite{ronetti_hong-ou-mandel_2019,battista_quantum_2013,PhysRevB.103.045427,PhysRevApplied.10.024007} and experimental \cite{pekola_towards_2015,RevModPhys.93.041001} contexts, as well as the references therein.  


We demonstrate that the heat Fano factor for tunneling between two gapless one-dimensional modes yields a universal number.
In particular, in the presence of a small temperature gradient between the two edges, $\Delta T \ll T$, the scaling dimension $h$ of the tunneling quasiparticle is immediately obtained from the heat Fano factor given in Eq.~\eqref{eq:Close_Temp}.
This property plays a crucial role in governing the edge dynamics of the gapless edge excitations~\cite{wen_quantum_2004}. For 
certain simple cases, such as
Abelian theories with no counter-propagating modes, the scaling dimension is directly related to the statistical phase obtained by braiding two quasiparticles, via $\theta= 2\pi h$. The scaling dimension can also be used to distinguish between underlying theories with otherwise similar quantum numbers, such as different candidate theories of the potentially non-Abelian $\nu=5/2$ FQH state~\cite{yang_influence_2013}. 

For completeness, we also obtain closed integral expressions for the heat Fano factor for two general edge temperatures, which solely depends on the scaling dimension. If $2h$ is an integer, and the temperature of the cold edge is zero, then this expression is reduced to fractions of Zeta functions (see App.~\ref{App:luttinger_liquid} and \ref{App:generic_cft} for more details).

\textit{Acknowledgements}-- We thank Francesco Buccheri, Bivas Dutta, Reinhold Egger, Edward Medina Guerra, Bo Han, Moty Heiblum, Abhay Nayak, Kyrylo Snizhko, Ady Stern and Gil Refael for useful discussions. This work was partially supported by grants from the ERC
under the European Union’s Horizon 2020 research and innovation programme
(grant agreements LEGOTOP No. 788715 and HQMAT No. 817799), the DFG
(CRC/Transregio 183, EI 519/7-1), the BSF and NSF (2018643), the ISF
Quantum Science and Technology (2074/19). N.S. was supported by the Clore Scholars Programme.\textbf{}
\bibliography{main}

\appendix
\begin{widetext}

\section{Scattering Theory}
    \label{App:scattering_theory}

\subsection{Non-interacting fermions and bosons}

We begin with calculations for non-interacting fermions and bosons, using the standard 
tool of scattering theory. Focusing on the geometry of Fig.~\ref{fig01}, We begin with the typical scattering theory formulas for the heat current and the
zero-frequency heat current auto-correlations (or noise) at terminal $\alpha$ \cite{blanter_shot_2000}. For heat current, these formulas are
\begin{align}
    \label{eq:canonical_scattering}
    \left\langle \hat{I}_{\alpha,E}\right\rangle  & =\frac{1}{2\pi\hbar}\sum_{\beta}\sum_{m}\int dE\left(E-\mu_{\alpha}\right)\left[\delta_{\alpha\beta}-\sum_{k}s_{\alpha\beta;mk}^{\dagger}\left(E\right)s_{\alpha\beta;km}\left(E\right)\right]f_{\beta}^{\pm}\left(E\right),\\
    \tilde{S}_{\alpha\alpha,E} & =\frac{1}{2\pi\hbar}\sum_{\gamma\delta}\sum_{nb}\int dE\left(E-\mu_{\alpha}\right)^{2}A_{\gamma\delta}^{mn}\left(\alpha;E,E\right)A_{\delta\gamma}^{nm}\left(\alpha;E,E\right)\times\\
     & \times\left\{ f_{\gamma}^{\pm}\left(E\right)\left[1\mp f_{\delta}^{\pm}\left(E\right)\right]+\left[1\mp f_{\gamma}^{\pm}\left(E\right)\right]f_{\delta}^{\pm}\left(E\right)\right\} .\nonumber 
\end{align}
Here, the $\pm$ refers to fermions/bosons, with $f_{\alpha}^{\pm}\left(E\right)=\left[\exp\left(\left(E-\mu_{\alpha}\right)/T_{\alpha}\right)\pm1\right]^{-1}$
the Fermi-Dirac/Bose-Einstein distribution function at terminal $\alpha$.
The tensor $A$ is given by
\begin{equation*}
A_{\beta\gamma}^{mn}\left(\alpha;E,E^{\prime}\right)=\delta_{mn}\delta_{\alpha\beta}\delta_{\alpha\gamma}-\sum_{k}s_{\alpha\beta;mk}^{\dagger}\left(E\right)s_{\alpha\gamma;kn}\left(E^{\prime}\right)
\end{equation*}
where $\alpha,\beta$ are terminals, $m,n$ are channels, $s_{\alpha\gamma;kn}\left(E^{\prime}\right)$
are scattering matrix components. In our geometry, there is only a single channel, and the scattering matrix is given by
\begin{equation*}
\left(\begin{array}{cccc}
0 & 0 & 1 & 0\\
0 & 0 & 0 & 1\\
t\left(E\right) & r\left(E\right) & 0 & 0\\
r\left(E\right) & t\left(E\right) & 0 & 0
\end{array}\right),
\end{equation*}
where we work in the basis $\left(\begin{array}{cccc}
S_{1} & S_{2} & D_{1} & D_{2}\end{array}\right)$. In other words, $r(E)$ is the amplitude associated with reflection between the two edges through the scatterer, and $t(E)$ is the amplitude associated with continuing along the original edge. Defining $\left|r(E)\right|^{2}\equiv R\left(E\right)$, charge conservation dictates that the scattering matrix is unitary, imposing~$\left|t(E)\right|^{2}\equiv1-R\left(E\right).$ We further
assume that all chemical potentials are zero, \textit{i.e.,} there is no biasing in the system, and that the source and drain on each edge are identical, $T_{D_{1}}=T_{S_{1}}\equiv T_{R};\;T_{D_{2}}=T_{S_{2}}\equiv T_{L}.$

We are interested in the heat current and the heat current fluctuations in the right-moving drain. Defining $I_E \equiv I_{D_1,E}$ and $\tilde{S}_E \equiv \tilde{S}_{D_1 D_1,E}$, this gives the explicit expression 
\begin{align}
    \label{eq:Scattering_QPC}
    I_{E} & =\frac{1}{h}\int dER\left(E\right)\left[f^{\pm}_{R}-f^{\pm}_{L}\right]E\\
    \tilde{S}_{E} & =\frac{2}{h} \int dE E^{2} \Big\{ \left( 2-R\left(E\right) \right) f^{\pm}_{R} \left[1\mp f^{\pm}_{R}\right] \\
     & \left.+R\left(E\right)f^{\pm}_{L} \left[1 \mp f^{\pm}_{L}\right]\pm R\left(E\right)\left(1-R\left(E\right)\right)\left(f^{\pm}_{R}-f^{\pm}_{L}\right)^{2}\right\} .\nonumber 
\end{align}

We solve these for a general scatterer obeying $R \left( E \right) = R_0 \left(E/E_c\right)^\alpha $, where $E_c \gg T_i$ is a high-energy cutoff. While a priori assuming a power-law dependence for the scatterer is an arbitrary choice, this is apt for the one-dimensional systems in question; Luttinger liquids in particular are notorious for power-law dependences of back-scattering from impurities \cite{kane_transport_1992,kane_transmission_1992}. As will be shown in the following sections, for a standard QPC, we expect $\alpha = 4 h -2$, where $h$ is the scaling dimension of the tunneling quasiparticle. This corresponds to $\alpha = 0$ for fermions (with $h=1/2$) and $\alpha = 2$ for bosons ($h=1$).

We wish to use these quantities to probe the quasiparticles which tunnel through the scatterer. As such, we focus only on the excess heat current fluctuations, \textit{i.e.,} the noise contribution that is obtained from a non-zero $R(E)$, and the tunneling heat current. We thus define the excess noise as, $S_E \equiv \tilde{S}_E-\tilde{S}_E \big|_{R(E)=0}$, and define our ``heat Fano factor" as $\mathcal{F}_E \equiv S_E/2 I_E$.

Calculating these integrals is now straight-forward. We focus on two particular limits of interest, to leading order in $R_0$. For nearby temperatures, $T_R \equiv T, T_L =T+\Delta T$, we obtain, for both fermions and bosons, Eq.~\eqref{eq:Scattering_Close} in the main text, with further corrections being of order $O \left( \frac{\Delta T}{T}, R_0 \right)$.

Conversely, when tunneling from a hot edge, $T_L = T$, to a cold edge, $T_R = 0$, we obtain
\begin{equation}
    \label{eq:Scattering_Zero}
    \mathcal{F}_E =
    \begin{cases}
    \left( \alpha + 2 \right) \frac{ \zeta \left(\alpha + 3 \right)}{ \zeta \left(\alpha + 2 \right) } k_B T + O \left(  R_0 \right) & \mathrm{bosons}\\
    \frac{2^{\alpha +2}-1}{2^{\alpha +1}-1} \frac{ \alpha + 2}{2} \frac{ \zeta \left(\alpha + 3 \right)}{\zeta \left(\alpha + 2 \right) } k_B T + O \left(  R_0 \right)& \mathrm{fermions}\\
    \end{cases}.
\end{equation}
Comparing these expressions with the generic-CFT results of Apps.~\ref{boson zero 3} and~\ref{app:CFT_zero_temp}, we indeed see that the two converge under $\alpha + 2 = 4 h_O$ for two cases: the bosonic case, in which $\alpha =2 \leftrightarrow h_0 = 1$; and the fermionic case, in which $\alpha = 0 \leftrightarrow h_0 = 1/2$.

The expert reader may be rightfully confused at the distinction we make between bosons and fermions, as we focus on one-dimensional systems, in which the two descriptions are physically equivalent. Indeed, this is the case, and we utilize the powerful tool of bosonization in App.~\ref{App:luttinger_liquid}, when we extend our results to Luttinger liquids. 

We emphasize, however, that by using scattering theory, we explicitly assume that the scattering of particles between the edges can be described using a bi-linear operator in terms of the creation/annihilation operators that are defined along the original edges. As such, for the results in this section, it is crucial that bosons may only scatter between edges that host non-interacting bosons, and fermions may only scatter between edges that host non-interacting fermions. It is thus inappropriate to describe boson scattering, for instance, using fermion edges and a ``bosonic" power law $\alpha = 2$.

This subtlety is also responsible for the discrepancy between Eq.~\eqref{eq:Scattering_Zero} and Eqs.~\eqref{eq:zero_temp_general_integer}, \eqref{eq:CFT_zero_temp_half_ints} and~\eqref{eq:CFT_zero_temp_ints} for half-integer scaling dimensions $h_O>1$. The derivation of an effective tunneling amplitude with a power law dependence $R(E) \sim E^{4h_O-2}$ explicitly requires a full calculation that accounts for the interacting nature of the system. As such, one cannot manually insert $R(E)\sim E^\alpha$ into an inherently non-interacting scattering theory and expect to recover known results with $\alpha = 4 h_O -2$. The fact that this ``naive" approach actually does work for the small-temperature gradient case, as can be seen by comparing Eq.~\eqref{eq:Scattering_Close} and Eq.~\eqref{eq:Close_Temp}, is rather remarkable. 

\subsection{Schematic measurement with quantum dot}
    \label{App:schematic_quantum_dot}
    
Here we schematically propose measurement of heat current fluctuations using a quantum dot. We present here scattering theory calculations for fermionic modes; any generalized approach to more sophisticated edges modes will require work of the nature explored in \cite{furusaki_resonant_1998}, and we leave this for future work. The goal here is mainly to demonstrate that the thermoelectric power generated by quantum dots leads to \textit{charge} current fluctuations of the carriers that tunnel through the dot, that encodes information about the \textit{heat} current fluctuations before it.

\begin{figure}
    \centering
    \includegraphics[width=0.6\columnwidth]{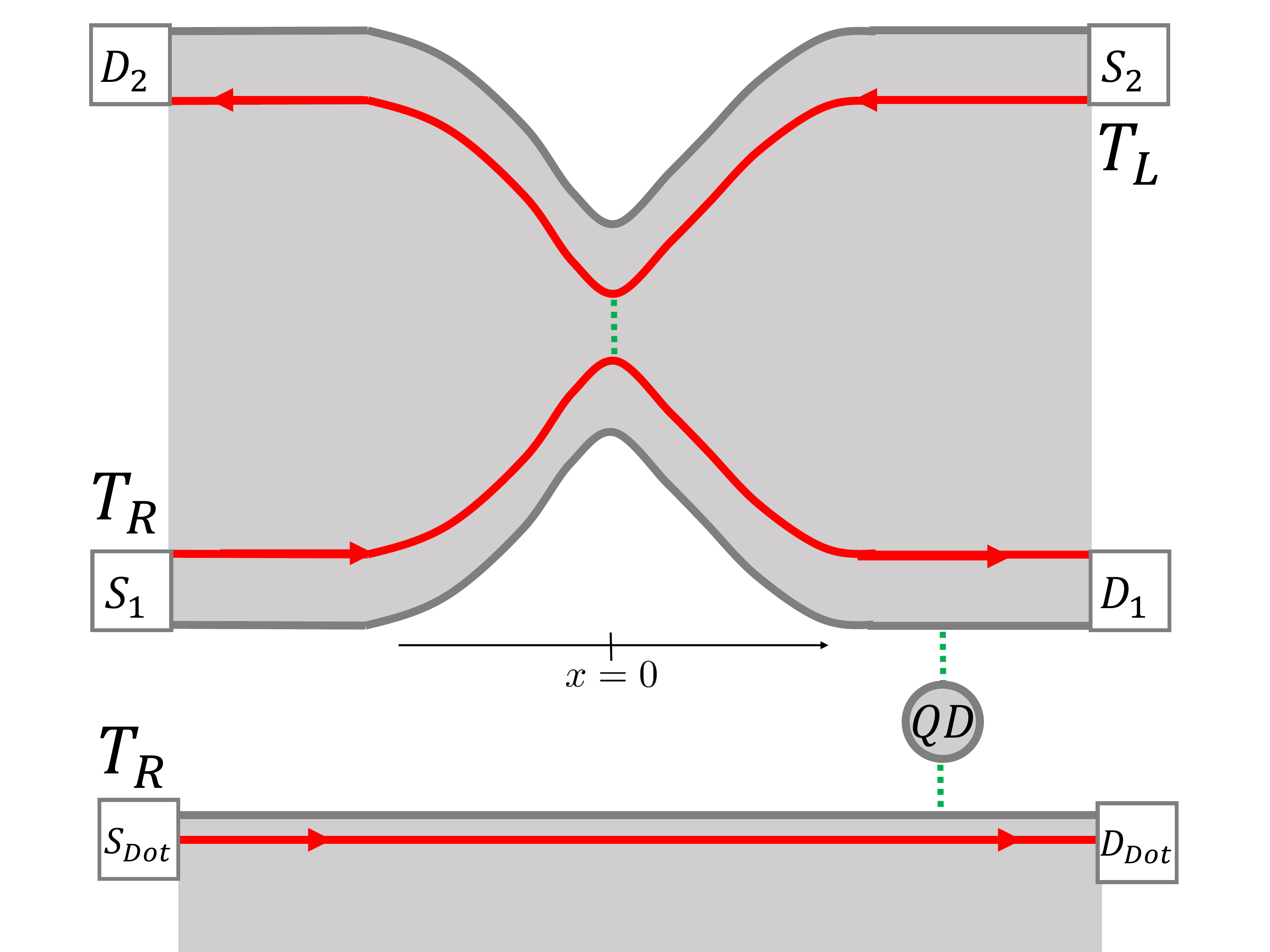}
    \caption{A configuration for measuring heat current fluctuations using a quantum dot. We take the original setup in Fig.~\ref{fig01}, and add a third edge state, which is coupled to the right-moving edge via a quantum dot. The thermoelectric effects of the quantum dot lead to \textit{charge} fluctuations of the carriers that tunnel through the dot and are gathered at the drain $D_{Dot}$. These are related to the \textit{heat} current fluctuations at the ``original" drain $D_1$.}
    \label{fig:QD_geometry}
\end{figure}

We assume a quantum dot is placed between the lower edge of our original system and a third, disconnected edge, as shown schematically in Fig.~\ref{fig:QD_geometry}. We treat the dot as having a single level, of energy $\varepsilon_0$, and width $\Gamma$, and assume there is no temperature gradient across the quantum dot, \textit{i.e.,} the disconnected edge is at temperature $T_R$. Similarly to the previous section, the scattering matrix of this entire system will be given by
\[
\left(\begin{array}{cccccc}
0 & 0 & 0 & 1 & 0 & 0\\
0 & 0 & 0 & 0 & 1 & 0\\
0 & 0 & 0 & 0 & 0 & 1\\
t_{\text{QPC}}t_{\text{D}} & r_{\text{QPC}}t_{\text{D}} & r_{\text{D}} & 0 & 0 & 0\\
r_{\text{QPC}} & t_{\text{QPC}} & 0 & 0 & 0 & 0\\
t_{\text{QPC}}r_{\text{D}} & r_{\text{QPC}}r_{\text{D}} & t_{\text{D}} & 0 & 0 & 0
\end{array}\right)
\]
where our basis is $\left(\begin{array}{cccccc}
S_{1} & S_{2} & S_{\text{Dot}} & D_{1} & D_{2} & D_{\text{Dot}}\end{array}\right).$

Plugging these terms into the canonical expressions for charge current and charge noise, we obtain, after the dot,

\begin{align}
    \left<I_{\text{Dot}} \right> & =\frac{e}{2\pi\hbar}\int dE\left|r_{\text{D}}\left(\varepsilon\right)\right|^{2}\left|r_{\text{QPC}}\left(\varepsilon\right)\right|^{2}\left(f_{1}\left(E\right)-f_{2}\left(E\right)\right)\\
    S_{\text{Dot,Dot}} & =\frac{e^{2}}{\pi\hbar}\int dE\left\{ 2f_{1}\left(E\right)\left[1 - f_{1}\left(E\right)\right]+\left|r_{\text{QPC}}\left(E\right)r_{\text{D}}\left(E\right)\right|^{2}\left(f_{2}\left(E\right)\left[1 - f_{2}\left(E\right)\right]-f_{1}\left(E\right)\left[1 - f_{1}\left(E\right)\right]\right)\right.\\
 & + \left|r_{\text{D}}\left(E\right)\right|^{2}\left|r_{\text{QPC}}\left(E\right)\right|^{2}\left(1-\left|r_{\text{D}}\left(E\right)\right|^{2}\left|r_{\text{QPC}}\left(E\right)\right|^{2}\right)\left(f_{1}(E)-f_{2}(E)\right)^{2}\nonumber 
\end{align}

We now use a standard reflection coefficient for the dot of $r\left(E\right)=i\frac{\Gamma}{\varepsilon-\varepsilon_{0}+i\Gamma},$ such that in the narrow-dot limit, $\Gamma \rightarrow 0$, we obtain $\left| r_D \left(E\right) \right|^2 = \Gamma \delta \left( E - \varepsilon_0 \right)$. We thus obtain
\begin{align*}
    \left< I_{\text{Dot}} \right> & =\frac{e}{2\pi\hbar}\Gamma R \left(\varepsilon_0 \right)\left(f_{1}\left(\varepsilon_{0}\right)-f_{2}\left(\varepsilon_{0}\right)\right)\\
    S_{\text{Dot,Dot}} & =4\frac{e^{2}}{h}k_{B}T_{1}+\Gamma R \left(\varepsilon_0 \right)\left(f_{2}\left(\varepsilon_{0}\right)\left[1- f_{2}\left(\varepsilon_{0}\right)\right]-f_{1}\left(\varepsilon_{0}\right)\left[1 - f_{1}\left(\varepsilon_{0}\right)\right]\right)\\
     & + \Gamma R \left(\varepsilon_0 \right)\left(f_{1}(\varepsilon_{0})-f_{2}(\varepsilon_{0})\right)^{2}.
\end{align*}
Subtracting the Johnson-Nyquist noise of $4\frac{e^{2}}{h}k_{B}T_{1}$, we see these expressions for the excess terms are identical to the integrands in Eq.~\eqref{eq:Scattering_QPC}, evaluated at $\varepsilon_0$, up to the replacements $R \left(E\right) \rightarrow \Gamma R \left(\varepsilon_0 \right)$ and $E/E^2 \rightarrow e/e^2$, showing a direct relationship between the heat fluctuations before the dot and the charge fluctuations after it.

\section{Luttinger Liquid}
    \label{App:luttinger_liquid}
This section is intended for proving Eq.~\eqref{eq:Close_Temp} in the case of Abelian FQH states with filling fraction $\nu$. In doing so, we illustrate several technical treatments, such as expansion of the correlators as a function of the temperature gradient~$\Delta T$, and contour integration, which are commonly used in the case of generic CFT approach in the subsequent section. 
 
We start with introducing the model. We consider edge modes of a FQH state with filling fraction $\nu$, and  introduce one quantum point contact (QPC) at coordinate $x=0$. For simplicity, we focus on Laughlin states, where $\nu=1/m$ for odd integer $m$, and note that results for more sophisticated FQH states are contained within the generic CFT case derived in App.~\ref{App:generic_cft}. We portray the configuration in Fig.~\ref{fig01} (the source and drain of the right-/left-moving edge mode are denoted as $S_{1/2}$ and $D_{1/2}$). Following Wen \cite{wen_quantum_2004}, we describe the low-energy excitations in terms of chiral boson fields, $\phi_{R/L}(x,t)=\phi_{R/L}(x\mp vt)$, satisfying the commutation relations $[\phi_{R/L}(x \mp vt_a),\phi_{R/L}(y \mp vt_b)]=\mp i\pi\Theta(x-y \mp v(t_{a}-t_b))$, with $\Theta$ being the Heaviside step function. The Hamiltonian is given by $ H=H_0+H_T $, where
\begin{subequations}
    \begin{align}
        H_0 & = \int_{-\infty}^{\infty}dx\frac{v}{4\pi}[(\partial_x\phi_R)^2+(\partial_x\phi_L)^2],\\
        H_T & = \Gamma_0 \Psi^{\dagger}_R(0) \Psi_L(0) + \text{h.c.}.\label{1}
    \end{align}
\end{subequations}
Here, $v$ is the velocity of the edge modes, $\Gamma_0$ is the amplitude associated with tunneling between the edges, and the operator $\Psi_{R/L}(x,t) \sim e^{i \sqrt{\nu}\phi_{R/L}(x,t)}$ annihilates a charge $e \nu$ quasiparticle on the right-/left-moving edge. We assume without loss of generality that $\Gamma_0$ is real. 
 
We study the heat current of the right-moving edge at the drain $D_1$, at coordinate $x=d~(>0)$, defined by $I_E=\frac{v^2}{4\pi}(\partial_x\phi_R(d,t))^2$. We
assume the amplitude of the tunneling $\Gamma_0$ is small, and expand the correction to the heat current up to the second order of $\Gamma_0$,
\begin{equation}
    I_E(t)= I_E^{(0)}(t)+I_E^{(1)}(t)+I_E^{(2)}(t)+\mathcal{O}(\Gamma_0^3),\label{IE}
\end{equation}
where
\begin{subequations}
    \begin{align}
         I_E^{(0)}(t) & =\frac{v^2}{4\pi}(\partial_x\phi_R(d,t))^2, \\
         I_E^{(1)}(t) & =i\int^t_{-\infty}dt^{\prime}[H_T(t^{\prime}),\frac{v^2}{4\pi}(\partial_x\phi_R(d,t))^2],\label{1st order0} \\
         I_E^{(2)}(t) & =i^2\int^t_{-\infty}dt^{\prime}\int^{t^{\prime}}_{-\infty}dt^{\prime\prime}\bigl[H_T(t^{\prime\prime}),[H_T(t^{\prime}),\frac{v^2}{4\pi}(\partial_x\phi_R(d,t))^2]\bigr].\label{2nd}
    \end{align}
\end{subequations}

We are interested in the current that tunnels between the edges, and hence are not interested in $I_E^{(0)}$. We begin calculating the tunneling current via Eq.~\eqref{1st order0}. Utilizing the bosonic commutation relations and the chiral nature of the modes, we find
\begin{eqnarray}
    I^{(1)}_E(t) = \Gamma_0\Psi_R^{\dagger}(\tilde{t})\{\partial_t\Psi_L(\tilde{t})\}+\Gamma_0\{\partial_t\Psi_L^{\dagger}(\tilde{t})\}\Psi_R(\tilde{t}).
    \label{1st}   
\end{eqnarray}
Here, we have introduced $\tilde{t}\equiv t-d/v$, and have suppressed the spatial coordinate $x=0$, defining $\Psi_{R/L}(t) \equiv \Psi_{R/L}(0,t)$, $\phi_{R/L}(t) \equiv \phi_{R/L}(0,t)$ for brevity. When we take the expectation value of $I^{(1)}_E(t)$ w.r.t. the ground state, it vanishes due to the ``neutrality condition" of the vertex operators, \textit{i.e.,} $\braket{I^{(1)}_E(t)}=0$. We remark this will be the case for any odd power of $\Gamma_0$.

We move on to the second order. Explicitly writing the terms $H_T$ in Eq.~\eqref{2nd}, utilizing the commutation relations, and taking the expectation value, we arrive after some algebra at the expression
\begin{eqnarray}
   \braket{I^{(2)}_E(t)}= 2i\Gamma_0^2\int^{\tilde{t}}_{-\infty}dt^{\prime\prime} 
\Bigl[G_R(t^{\prime\prime}-\tilde{t})\{\partial_tG_L(t^{\prime\prime}-\tilde{t})\}
    +
     G_R(\tilde{t}-t^{\prime\prime})\{\partial_tG_L(\tilde{t}-t^{\prime\prime})\}\Bigr], \label{2nd2}
\end{eqnarray}
where we have introduced the correlation function
\begin{eqnarray}
       G_{R/L}(t_a-t_b)&=&\braket{\Psi_{R/L}^{\dagger}(t_a)\Psi_{R/L}(t_b)}=\braket{\Psi_{R/L}(t_a)\Psi_{R/L}^{\dagger}(t_b)}\nonumber\\
       &=&\Biggl[\frac{(\pi T_{R/L}/(i\Lambda))}{\sinh{\pi T_{R/L}((t_a-t_b)-i\epsilon)}}\Biggr]^{\nu}.\label{GR}
\end{eqnarray}
Here, $\Lambda$ is a UV cutoff, $\epsilon>0$ is a short time cutoff, and $T_{R/L}$ is the temperature of the right-/left-moving edge mode. We note that the second equation comes from the fact that, in the absence of a bias voltage, particle-hole symmetry is respected. 
Changing variables as $\tau=t^{\prime\prime}-\tilde{t}$, and using the relation $\partial_tG_R(\tilde{t}-t^{\prime\prime})=-\partial_{t^{\prime\prime}}G_R(\tilde{t}-t^{\prime\prime})$, we rewrite Eq.~\eqref{2nd2} as
\begin{equation}
    \boxed{
  \braket{I^{(2)}_E(t)}=-2i\Gamma_0^2\int_{-\infty}^{+\infty}d\tau G_R(\tau)\partial_{\tau}G_L(\tau)}. \label{th}
\end{equation}

Note that when the temperature of the edges are equal, \textit{i.e.,} $T_R=T_L$, the correlators $G_R(\tau)$ and $G_L(\tau)$ are identical. Eq.~\eqref{th} then vanishes, as the integrand is proportional to total derivative of $G_L^2(\tau)$. This is physically plausible since the geometry we consider becomes symmetric when $T_R=T_L$ and the amount of heat transferred from the right-moving edge to the left-moving edge through the QPC is precisely compensated by the one from the left-moving edge to the right-moving edge, leading to zero net heat transfer. 

In the following, we study the correction of the heat current Eq.~\eqref{th} in two different limits: one is the case where we introduce an infinitesimally small temperature gradient between the right- and left-moving edges; and the other is where we set the temperature of one edge to be zero. 

\subsection{Almost equal temperature limit }\label{app expansion}
As we mentioned previously, when the temperature of the both edges are equal, \textit{i.e.,} $T_R=T_L$, the second order correction in Eq.~\eqref{th} vanishes. We slightly deviate from this limit, setting the temperature of the left-moving edge as $T_L=T+\Delta T$ ($|\Delta T/T|\ll 1$) while keeping $T_R=T$. From Eq.~\eqref{GR}, we have
\begin{equation}
     \partial_{\tau}G_R(\tau)=-\nu(\pi T_R)\coth(\pi T_R(\tau-i\epsilon))G_R(\tau),\label{derivative}
\end{equation}
allowing us to rewrite Eq.~\eqref{th} as
\begin{equation*}
      \braket{I^{(2)}_E(t)} =
      \frac{2\Gamma_0^2\nu(\pi T_R)^{\nu+1}(\pi T_L)^{\nu}}{\Lambda^{2\nu}}\int^{+\infty}_{-\infty}d\tau\frac{\cosh(\pi T_R(\tau-i\epsilon))}{[i\sinh(\pi T_R(\tau-i\epsilon))]^{\nu+1}[i\sinh(\pi T_L(\tau-i\epsilon))]^{\nu}}.
\end{equation*}
Changing variables as $\lambda=\frac{T_L}{T_R}=1+\frac{\Delta T}{T}$, $z=\pi T_R(\tau-i\epsilon)+\frac{i \pi}{2}$, it can be further transformed into
\begin{equation*}
  \braket{I^{(2)}_E(t)}=\frac{2\Gamma_0^2\nu(\pi T)^{2\nu}\lambda^{\nu}}{\Lambda^{2\nu}}\int^{+\infty+\frac{i \pi}{2}-i\epsilon}_{-\infty+\frac{i \pi}{2}-i\epsilon}dz\frac{-i\sinh z}{[\cosh z]^{\nu+1}[i\sinh(\lambda(z-\frac{i \pi}{2})]^{\nu}}. \end{equation*}
Since the integrand does not have any pole between the real axis and $z=i\pi/2$, one can shift the integration on the complex plane back to the real axis, obtaining
\begin{equation}
  \braket{I^{(2)}_E(t)}=\frac{2\Gamma_0^2\nu(\pi T)^{2\nu}\lambda^{\nu}}{\Lambda^{2\nu}}\int^{+\infty}_{-\infty}dz\frac{-i\sinh z}{[\cosh z]^{\nu+1}[i\sinh(\lambda(z-\frac{i \pi}{2})]^{\nu}}. \label{equal}
  \end{equation}

As we assume $|\Delta T/T|\ll 1$, we expand Eq.~\eqref{equal} in powers of $\Delta T/T$. We obtain 
\begin{subequations}
    \begin{align}
       \braket{I^{(2)}_E(t)} & \simeq -i\frac{2\Gamma_0^2\nu(\pi T)^{2\nu}}{\Lambda^{2\nu}}\int^{+\infty}_{-\infty}dz\frac{\sinh z(1+\nu \frac{\Delta T}{T})\Bigl(1-\nu(z-i\frac{\pi}{2})\tanh(z)\frac{\Delta T}{T}\Bigr)}{[\cosh z]^{2\nu+1}} \\ \label{expansion}
       & =  -i\frac{2\Gamma_0^2\nu(\pi T)^{2\nu}}{\Lambda^{2\nu}}\int^{+\infty}_{-\infty}dz\frac{\sinh z}{[\cosh z]^{2\nu+1}} \\
       & + \frac{\pi\Gamma_0^2\nu^2(\pi T)^{2\nu}}{\Lambda^{2\nu}}\int^{+\infty}_{-\infty}dz\Biggl(\frac{1}{[\cosh z]^{2\nu}}-\frac{1}{[\cosh z]^{2\nu+2}}\Biggr)\frac{\Delta T}{T} + \mathcal{O} \left( \left( \frac{\Delta T}{T}\right)^2\right).
    \end{align}
\end{subequations}
The $\mathcal{O}(1)$ term is zero as the integrand is odd function of $z$, which is consistent with that fact that  $\braket{I^{(2)}_E(t)}$ vanishes at equal temperature, $T_R=T_L$. The integrals comprising the first order, $\mathcal{O} \left(\frac{\Delta T}{T} \right)$ term can be described using the Euler Gamma function, giving us the final expression
\begin{equation}
    \boxed{ \braket{I^{(2)}_E(t)}_{\substack{T_L=T+\Delta T \\ T_R=T}}=\frac{\pi\Gamma_0^2(\pi T)^{2\nu}}{\Lambda^{2\nu}}\frac{2^{2\nu-1}}{2\nu+1}\frac{\Gamma^2(\nu+1)}{\Gamma(2\nu)}\frac{\Delta T}{T}+\mathcal{O}\Bigl(\biggl(\frac{\Delta T}{T}\biggr)^3\Bigr)}.\label{current1}
\end{equation}
Based on this result, we find the amount of the heat current gets increased when the sign of $\Delta T$ is positive, which is consistent with the physical intuition that the heat flows from the hotter edge (left-moving edge) to the cold edge (right-moving edge). The same intuition leads one to surmise that all even orders in $\frac{\Delta T}{T}$ will vanish, as the total tunneling current must be odd in the temperature gradient. 
 
It is interesting to note that 
when $\nu=1/2$, which corresponds to a spin liquid phase known as the Kalmeyer-Laughlin state~\cite{KLstate1987}, the heat current is proportional to $\Delta T$, not depending on~$T$. 

\subsection{Zero temperature limit}\label{boson zero 2}
We take a different limit by setting the temperature of the right-moving edge to be zero and the one of the left-moving edge to be $T$, \textit{i.e.,} $T_R=0$, $T_L=T$. 
In this limit, the correlator~$G_R(\tau)$ defined in Eq.~\eqref{GR} changes as $G_R(\tau) \rightarrow \Bigl[i\Lambda(\tau-i\epsilon)\Bigr]^{-\nu}$.
The heat current of Eq.~\eqref{th} becomes
\begin{equation}
   \braket{I^{(2)}_E(t)}
 =\frac{2\Gamma_0^2\nu(\pi T)^{\nu}}{\Lambda^{2\nu}} \int_{-\infty}^{+\infty}d\tau \frac{1}{[i\sinh(\pi T(\tau-i\epsilon))]^{\nu}}
 \Biggl[\frac{1}{i(\tau-i\epsilon)}\Biggr]^{\nu+1}.\label{21}
\end{equation}
Changing variables as $z=\pi T(\tau-i\epsilon)+\frac{i\pi}{2}$, a similar line of argument to Eq.~\eqref{equal} shows that Eq.~\eqref{21} is transformed into
\begin{equation}
    \boxed{
  \braket{I^{(2)}_E(t)}_{\substack{T_L=T\\ T_R=0}}= \frac{2\Gamma_0^2}{\Lambda^{2\nu}} \int_{-\infty}^{+\infty }dz\frac{\nu (\pi T)^{2\nu}}{[\cosh(z)]^{\nu}[i(z-i\pi/2)]^{\nu+1}}} .\label{23}
\end{equation}
 \subsubsection{A special case: fermionic hopping}
It is challenging to estimate Eq.~\eqref{23} analytically for generic values of $\nu$. However, we can evaluate Eq.~\eqref{23} explicitly in the case of $\nu$ being an integer. As a simple example, when $\nu=1$, Eq.~\eqref{23} becomes
\begin{equation}
    \braket{I^{(2)}_E(t)}= \frac{2\Gamma_0^2 (\pi T)^{2}}{\Lambda^{2}} \int_{-\infty}^{+\infty }dz\frac{1}{[\cosh(z)][i(z-i\pi/2)]^{2}}.\label{24}
\end{equation}
\begin{figure}[h] 
        \centering
        \includegraphics[width=0.25\columnwidth]{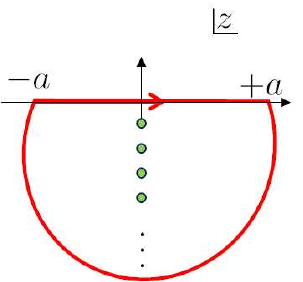}
        \caption{The contour of the integration is portrayed by red curves. This contour encloses poles of the first order, $z_n=i\pi/2-in\pi (n\in\mathbb{N})$ illustrated by green dots.} \label{fig2} 
\end{figure}
To perform the integral, we introduce a semi-circle contour in the lower half complex plane with a radius of $a$, as depicted in Fig.~\ref{fig2}, and take the limit $a\to+\infty$. Denoting this contour as $C$, the integral in Eq~\eqref{24} is identical to $\oint_Cdz F(z)$ ($F(z)=\frac{1}{[\cosh(z)][i(z-i\pi/2)]^{2}}$) as the integration along the curved line away from the real axis becomes zero in the limit $a\to+\infty$. The contour $C$ encloses poles at $z_n=i\pi/2-in\pi$ $(n\in\mathbb{N})$, all of which are order of one, with a residue of $\frac{(-1)^n}{i(n\pi)^2}$. We can then calculate the integral by using the Cauchy theorem, giving
\begin{equation*}
    \braket{I^{(2)}_E(t)}=- \frac{4 \pi T^2 \Gamma_0^2}{\Lambda^{2}} \sum_{n=1}^{\infty}\frac{(-1)^n}{n^2}.
\end{equation*}
Now using the identity $\sum_{n=1}^{\infty}\frac{(-1)^n}{n^2}=-\frac{1}{2}\zeta(2)=-\frac{\pi^2}{12}$,
where $\zeta(s)$ is the Riemann zeta function, we get
\begin{equation}
   \boxed{ \braket{I^{(2)}_E(t)}_{\nu=1}= \frac{2\Gamma_0^2 }{\Lambda^{2}}\pi T^2 \zeta(2)=\frac{\pi^3 \Gamma_0^2 T^{2}}{3\Lambda^{2}} }.\label{zero1}
\end{equation}
\subsubsection{Generic integer case}\label{ap generic abelian}
We can evaluate Eq.~\eqref{23} for generic integer $\nu$, corresponding to multi-channel electron tunnelling at the QPC, by implementing the same contour integration. In doing so, we encounter a complication of evaluating the residue when we consider integer values $\nu>2$. To highlight this issue, let us focus on the case with $\nu=3$. In the vicinity of the pole $z_n=-in\pi+i\pi/2$, we expand $\frac{1}{[\cosh z]^3}$ as
\begin{equation*}
      \frac{1}{[\cosh(z)]^3}\simeq\frac{1}{-i(-1)^n(z-z_n)^3}[1-\frac{1}{2!}(z-z_n)^2+\cdots]=\frac{1}{-i(-1)^n}\biggl[\frac{1}{(z-z_n)^3}-\frac{1}{2(z-z_n)}\biggr]+\cdots,
\end{equation*}
indicating $\frac{1}{[\cosh z]^3}$ has multiple non-zero Laurent coefficients of negative powers.
If we go to the case larger integer values of $\nu$, the number of non-zero Laurent coefficients is increasing.
Indeed, for the generic integer of $\nu$, we expand $1/[\cosh z]^\nu$ as 
\begin{equation}
   \frac{1}{[\cosh(z)]^\nu}\simeq\frac{1}{i^\nu(-1)^{n\nu}}\biggl[\frac{1}{(z-z_n)^\nu}+\sum_{k=1}^{\infty}\binom{\nu+k-1}{k}(-1)^k \sum_{\substack{\{m_{l}\}\geq1\\1\leq l\leq k\\
        2\sum_lm_l<\nu
        }}
   \frac{(z-z_n)^{2\sum_lm_l-\nu}}
   {\prod_{l}(2m_l+1)!}
   \biggr]+\cdots. \label{poles2}
\end{equation}
Calculating the residues of the integrand in Eq.~\eqref{23} by making use of the expansion Eq.~\eqref{poles2}, we obtain the expression for the second order of the heat current. 
For odd integer values of $\nu$, we have
\begin{subequations}
    \begin{align}
        \braket{I^{(2)}_E(t)}&=\frac{2\Gamma_0^2}{\Lambda^{2\nu}}\nu(\pi T)^{2\nu}\sum_{0\leq a\leq \frac{\nu-1}{2}}p_a\eta(\nu+1+2a), \\
        p_{\frac{\nu-1}{2}}&=\frac{2}{\pi^{2\nu-1}}\binom{2\nu-1}{\nu} \\
        p_a&=\frac{2}{\pi^{\nu+2a}}\binom{\nu+2a}{2a}\sum_{\substack{k\geq 1\\}} 
        \sum_{\substack{\{m_{l}\}\geq1\\1\leq l\leq k\\
        2\sum_lm_l=\nu+1-2a
        }}
        \binom{\nu+k-1}{k}\frac{(-1)^{k+a+\frac{\nu-1}{2}}}{\prod_{l}(2m_{l}+1)!}
        \;(0\leq a \leq \frac{\nu-1}{2}-1),\label{b40}
    \end{align}
\end{subequations}
where $\eta(s)$ is the Dirichlet eta function, $\eta(s)=\sum_{n=1}^{\infty}\frac{(-1)^{n+1}}{n^s}=(1-2^{1-s})\zeta(s)$.
For even integer values of $\nu$, one finds
\begin{subequations}
    \begin{align}
        \braket{I^{(2)}_E(t)}&=\frac{2\Gamma_0^2}{\Lambda^{2\nu}}\nu(\pi T)^{2\nu}\sum_{1\leq b\leq \frac{\nu-1}{2}}q_b\zeta(\nu+2b) \\
         q_{\nu/2}&=\frac{2}{\pi^{2\nu-1}}\binom{2\nu-1}{\nu} \\
        q_b&=\frac{2}{\pi^{\nu+2b-1}}\binom{\nu+2b-1}{2b}\sum_{\substack{k\geq 1\\}} 
        \sum_{\substack{\{m_{l}\}>0\\1\leq l\leq k\\
        2\sum_lm_l=\nu-2b
        }}
        \binom{\nu+k-1}{k}\frac{(-1)^{k+b+\frac{\nu}{2}}}{\prod_{l}(2m_{l}+1)!}\;\;(1\leq b \leq \frac{\nu}{2}-1)\label{b41}
    \end{align}
\end{subequations}

\subsection{Heat current fluctuations}
After calculating the heat current operators, we will study the heat current fluctuations, given by the auto-correlations of heat current at the drain $D_1$ with coordinate $x=d>0$. Particularly, as we mentioned earlier, we are interested in the ratio between the heat current fluctuations and the heat current. 
To this end, we define the heat current fluctuations as
\begin{align*}
      \tilde{S}_E(\omega) & = \int_{-\infty}^{+\infty}dt_{12}e^{i\omega t_{12}} \tilde{S}_E(t_{12}) \\
       \tilde{S}_E(t_{12})  \equiv \braket{\{\Delta I_E(t_1),\Delta I_E(t_2)\}} & =
    \braket{\{I_E(t_1),I_E(t_2)\}}-2\braket{I_E(t_1)}\braket{I_E(t_2)}
\end{align*}
where $t_{12}=t_1-t_2$. Similarly to the heat current, we expand the heat current fluctuations up to the second order of the tunneling amplitude, $\Gamma_0$. The first order vanishes due to the neutrality condition, analogously to the previous argument on the heat current. Regarding the second order, there are three terms, namely,
\begin{subequations}
    \begin{align}
        \tilde{S}_E(t_{12}) & = S_E^{(00)}(t_{12})+S_E^{(11)}(t_{12})+S_E^{(02)}(t_{12})+S_E^{(20)}(t_{12})+\mathcal{O}(\Gamma_0^3), \label{nl} \\
        S_E^{(ij)}(t_{12}) & =\braket{\{\Delta I^{(i)}_E(t_1),\Delta I^{(j)}_E(t_2)\}}
        =\braket{\{I_E^{(i)}(t_1),I_E^{(j)}(t_2)\}}-2\braket{I_E^{(i)}(t_1)}\braket{I_E^{(j)}(t_2)} \label{0cftss}, \;(i,j=0,1,2).
    \end{align}
\end{subequations}
Notice that due to the charge-neutrality condition, $S_E^{(01)}(t_{12})$ and $S_E^{(10)}(t_{12})$ are zero. We are interested in the \textit{excess} heat current fluctuations~$S_E(t_{12})$, defined by subtracting the unperturbed fluctuations from the total fluctuations, \textit{i.e.,} $S_E(t_{12})= \tilde{S}_E(t_{12})-S_E^{(00)}(t_{12})$. 
Below, we will intently study the second order corrections of the heat current fluctuations, corresponding to the last three terms in Eq.~\eqref{nl}, $S_E^{(11)}(t_{12})$, $S_E^{(02)}(t_{12})$, and $S_E^{(20)}(t_{12})$.

\subsection{The auto-correlation term}
\label{app:auto_boson}
We first evaluate the term $S_E^{(11)}(t_{12})$, which corresponds to auto-correlations of the tunneling current. Referring to Eq.~\eqref{1st}, and noting $\braket{I_E^{(1)}(t_{1/2})}=0$, one finds
\begin{eqnarray*}
   S_E^{(11)}(t_{12})  =2\Gamma_0^2G_R(t_{12})
\{\partial_{t_1}\partial_{t_2}G_L(t_{12})\}+2\Gamma_0^2G_R(-t_{12})
\{\partial_{t_1}\partial_{t_2}G_L(-t_{12})\}.
\end{eqnarray*}
Implementing a Fourier transformation of this yields
\begin{equation}
    \boxed{
S_E^{(11)}(\omega)= -4\Gamma_0^2\int^{+\infty}_{-\infty}d\tau \cos(\omega\tau)G_R(\tau)\partial^2_{\tau}G_{L}(\tau).}\label{11term}
\end{equation}
\subsection{The cross terms}
We move on to evaluation of the ``cross terms", $S_E^{(02)}(t_{12})$, $S_E^{(20)}(t_{12})$, which represent the correlations of the tunneling current with the unperturbed current.
Referring to Eqs.~\eqref{IE},\eqref{2nd}, one finds
\begin{equation}
    \begin{aligned}
        \braket{I_E^{(0)}(t_1)I_E^{(2)}(t_2)}
        =\frac{i\Gamma_0^2}{2\pi}\int^{\tilde{t}_2}_{-\infty}dt^{\prime\prime} \braket{(\partial_{t_1}\phi_R(\tilde{t}_1))^2\Psi_R^{\dagger}(t^{\prime\prime})\Psi_R(\tilde{t}_2)}\partial_{t_2}G_L(t^{\prime\prime}-\tilde{t}_2) \\
        -i\frac{\Gamma_0^2}{2\pi}\int^{\tilde{t}_2}_{-\infty}dt^{\prime\prime}\braket{(\partial_{t_1}\phi_R(\tilde{t}_1))^2\Psi_R(\tilde{t}_2)\Psi_R^{\dagger}(t^{\prime\prime})}\partial_{t_2}G_L(\tilde{t}_2-t^{\prime\prime}),\label{02}
     \end{aligned}
\end{equation}
where we have introduced the abbreviated notation $\tilde{t}_{1/2}=t_{1/2}-d/v$.

To evaluate the correlation function involving the right-moving fields in Eq.~\eqref{02}, we exploit the following formulas
\begin{equation}
    \label{b formula}
    \braket{(\partial_{\tau_1}\phi_R(\tau_1))^2e^{\pm i\sqrt{\nu}\phi_R(\tau_3)}e^{\mp i\sqrt{\nu}\phi_R(\tau_4)}} = [(\partial_{\tau_1}\phi_R(\tau_1))^2-\nu K^2(\tau_1,\tau_3,\tau_4)]G_R(\tau_3-\tau_4) 
\end{equation}
where we define 
\begin{equation*}
    K(\tau_1,\tau_3,\tau_4)\equiv \braket{\partial_{\tau_1}\phi_R(\tau_1)\phi_R(\tau_3)}-\braket{\partial_{\tau_1}\phi_R(\tau_1)\phi_R(\tau_4)}.
\end{equation*}
This formula is obtained by introducing the correlator,
\begin{equation*}
    \begin{aligned}
        P(\{\epsilon_i\},\{\tau_i\}) & \equiv \braket{e^{-i\epsilon_1\phi_R(\tau_1)}e^{-i\epsilon_2\phi_R(\tau_2)}e^{-i\epsilon_3\phi_R(\tau_3)}e^{-i\epsilon_4\phi_R(\tau_4)}} \\ \label{boson P}
        & = \exp\biggr[-\frac{1}{2}\sum_i\epsilon_i^2\braket{\phi_R^2(\tau_i)}-\sum_{i<j}\epsilon_i\epsilon_j\braket{\phi_R(\tau_i)\phi_R(\tau_j)}\biggr], 
    \end{aligned}
\end{equation*}
and utilizing the relation
\begin{equation*}
      \braket{(\partial_{\tau_1}\phi_R(\tau_1))^2e^{\pm i\sqrt{\nu}\phi_R(\tau_3)}e^{\mp i\sqrt{\nu}\phi_R(\tau_4)}}=-\partial_{\tau_1}\partial_{\tau_2}\biggl[\lim_{\epsilon_1,\epsilon_2\to 0}\partial_{\epsilon_1}\partial_{\epsilon_2}P(\{\epsilon_i\},\{\tau_i\})\biggr]\Bigg|_{\substack{\tau_1=\tau_2\\ \epsilon_3=-\epsilon_4=\sqrt{\nu}}}.
\end{equation*}

Turning back to the correlators in Eq.~\eqref{02}, and making use of the formulas Eq.~\eqref{b formula}, one finds

\begin{equation}
    \begin{aligned}
    \braket{I_E^{(0)}(t_1)I_E^{(2)}(t_2)}&=&\frac{i\Gamma_0^2}{2\pi}\int^{\tilde{t}_2}_{-\infty}dt^{\prime\prime} \bigl[\braket{(\partial_{t_1}\phi_R(\tilde{t}_1))^2}-\nu K^2(\tilde{t}_1,t^{\prime\prime},\tilde{t}_2)\bigr]G_R(t^{\prime\prime}-\tilde{t}_2)\partial_{t_2}
    G_L(t^{\prime\prime}-\tilde{t}_2) \\
    &-&\frac{i\Gamma_0^2}{2\pi}\int^{\tilde{t}_2}_{-\infty}dt^{\prime\prime} \bigl[\braket{(\partial_{t_1}\phi_R(\tilde{t}_1))^2}-\nu K^2(\tilde{t}_1,t^{\prime\prime},\tilde{t}_2)\bigr]G_R(\tilde{t}_2-t^{\prime\prime})\partial_{t_2}G_L(\tilde{t}_2-t^{\prime\prime}).\label{ppq}
    \end{aligned}
\end{equation}
Using the standard free boson correlator, $\braket{\phi_R(t)\phi_R(0)}=-\ln{ \left[\frac{\sinh(\pi T_R(t-i\epsilon))}{(\pi T_R/i\Lambda)}\right]}$, the term $K(\tilde{t}_1,t^{\prime\prime},\tilde{t}_2)$ is rewritten as
\begin{equation}
    K^2(\tilde{t}_1,t^{\prime\prime},\tilde{t}_2) = \frac{(\pi T_R)^2\sinh^2{(\pi T_R(t^{\prime\prime}-\tilde{t}_2))}}{\sinh^2{(\pi T_R(\tilde{t}_1-\tilde{t}_2-i\epsilon))}\sinh^2(\pi T_R(\tilde{t}_1-t^{\prime\prime}-i\epsilon))}.\label{large K}
\end{equation}
By similar arguments, we find that the correlator $\braket{I_E^{(2)}(t_2)I_E^{(0)}(t_1)}$ has the same integral expression as Eq.~\eqref{ppq}. Therefore, $ \braket{\{I_E^{(0)}(t_1),I_E^{(2)}(t_2)\}}= 2\braket{I_E^{(0)}(t_1)I_E^{(2)}(t_2)}$.

By changing variables $\tau=t^{\prime\prime}-\tilde{t}_2$, we note that the first and third term in Eq.~\eqref{ppq} are summed up to be 
\begin{eqnarray*}
-\frac{iv^2\Gamma_0^2}{\pi}\braket{(\partial_{t_1}\phi_R(\tilde{t}_1))^2}\int^{+\infty}_{-\infty}d\tau G_R(\tau)\partial_{\tau}G_L(\tau).
\end{eqnarray*}
This is equivalent to $2\braket{I_E^{(0)}(t_1)}\braket{I_E^{(2)}(t_2)}$, which is canceled by the second term in Eq.~\eqref{0cftss}.
Hence, we are left with
\begin{eqnarray*}
    S_E^{(02)}(t_{12})&=& -
    \frac{i\Gamma_0^2\nu}{\pi }\int^{\tilde{t}_2}_{-\infty}dt^{\prime\prime} [K^2(\tilde{t}_1,t^{\prime\prime},\tilde{t}_2)G_R(t^{\prime\prime}-\tilde{t}_2)]\partial_{\tilde{t}_2}G_L(t^{\prime\prime}-\tilde{t}_2) \\
    &+& \frac{i\Gamma_0^2\nu}{\pi }\int^{\tilde{t}_2}_{-\infty}dt^{\prime\prime} [K^2(\tilde{t}_1,t^{\prime\prime},\tilde{t}_2)G_R(\tilde{t}_2-t^{\prime\prime})]\partial_{\tilde{t}_2}G_L(\tilde{t}_2-t^{\prime\prime}).
\end{eqnarray*}

We now perform a Fourier transformation of the cross terms.
Focusing on the heat current fluctuations $S_E^{(02)}(\omega)$, we find
\begin{subequations}
    \begin{align}
    S_E^{(02)}(\omega) &= \frac{i\Gamma_0^2\nu}{\pi }\int^{+\infty}_{-\infty}dt_{12}\int^{0}_{-\infty}d\tau e^{i\omega t_{12}} K^2(t_{12},\tau)G_R\partial_{\tau}G_L\label{cross1}\\
    &+\frac{i\Gamma_0^2\nu}{\pi }\int^{+\infty}_{-\infty}dt_{12}\int^{+\infty}_{0}d\tau e^{i\omega t_{12}} K^2(t_{12},-\tau)G_R\partial_{\tau}G_L.\label{cross term}
    \end{align}
\end{subequations}
Here, we changed variables to $\tau=t^{\prime\prime}-\tilde{t}_2$, and use an abbreviated expression as $G_{R/L}=G_{R/L}(\tau)$. Due to the change of variables, $  K^2(\tilde{t}_1,t^{\prime\prime},\tilde{t}_2)$ given in Eq.~\eqref{large K} has only $t_{12}$ and $\tau$ dependence. We see the integration over $t_{12}$ affects only the term proportional to $K^2(t_{12},\tau)$; we perform this integral via contour integration. Introducing the upper half-plane contour in the complex plane and resorting to the Cauchy theorem, it can be shown that
\begin{align*}
    \int^{+\infty}_{-\infty}dt_{12} e^{i\omega t_{12}} K^2(t_{12},\tau) &=2\pi i \sum_{n=0}^{\infty}\Bigl\{e^{-\frac{n\omega}{T_R}}i\omega(1+e^{i\omega \tau})+2e^{-\frac{n\omega}{T_R}}(1-e^{i\omega\tau})(\pi T_R)\coth(\pi T_R\tau)\Bigr\} \\
    &=i\pi N(\omega)\Bigl\{i\omega(1+e^{i\omega \tau})+2(1-e^{i\omega\tau})(\pi T_R)\coth(\pi T_R\tau)\Bigr\}. 
\end{align*}
where in the second line we've utilized the convergence of the infinite sum, and defined $N \left( \omega \right) \equiv \coth \left( \omega/2T_R \right)+1$. Using this and Eq.~\eqref{derivative}, we can rewrite the two terms comprising $S_E^{(02)}(\omega)$ as 

\begin{eqnarray*}
 \eqref{cross1} &=& -\Gamma_0^2N(\omega)\int_{-\infty}^0d\tau[i\omega\nu(1+e^{i\omega\tau})G_R-2(1-e^{i\omega\tau})\partial_{\tau}G_R]\partial_{\tau}G_L \\
 \eqref{cross term} &=& -\Gamma_0^2N(\omega)\int_{0}^{+\infty}d\tau[i\omega\nu(1+e^{-i\omega\tau})G_R+2(1-e^{-i\omega\tau})\partial_{\tau}G_R]\partial_{\tau}G_L.
\end{eqnarray*}

The second cross term, $S_E^{(20)}(\omega)$, can be approached in a similar manner. A direct calculation shows that $S_E^{(02)}(\omega)/N(\omega)=-S_E^{(20)}(-\omega)/N(-\omega)$.
Conversely, this can be derived via the Kubo-Martin-Schwinger relation~\cite{kubo1957,PhysRev.115.1342}; indeed, from this relation, we have
\begin{equation}
      S_E^{(20)}(\omega)=S_E^{(02)}(-\omega)e^{\frac{\omega}{T_R}},\label{kms}
\end{equation}
which follows directly from the relation $N(-\omega) e^{\frac{\omega}{T_R}} = -N(\omega)$.

In total, the sum of both cross terms is now explicitly written as
\begin{align}
    \label{bsim}
    {S}^{(02)}(\omega)+{S}^{(20)}(\omega) &= -2N(\omega)\Gamma_0^2 \int^{+\infty}_{-\infty} d\tau
    \left[ i\omega\nu \left(1 + \cos \left(\omega\tau\right) \right) G_R \partial_{\tau} G_L + 2i \sin \left(\omega\tau\right) \partial_{\tau}G_R\partial_{\tau}G_L)\right] \\ 
    & = -2N(\omega)\Gamma_0^2\int^{+\infty}_{-\infty}d\tau
    \left[i\omega\nu \left(1+\cos \left(\omega\tau\right) \right) G_R\partial_{\tau}G_L-2i\omega\cos \left(\omega\tau\right) G_R\partial_{\tau}G_L-2i\sin \left(\omega\tau\right) G_R\partial^2_{\tau}G_L \right] \nonumber 
\end{align}
where the second line is obtained by performing integration by parts on the second term.

Now we are in a good place to take the DC limit, $\omega\to0$. In this limit, $N(\omega)\omega\to 2T_R$. Recalling the second order correction of the heat current, Eq.~\eqref{th}, the first two terms of Eq.~\eqref{bsim} become
\begin{equation*}
    4T_R(\nu-1)\braket{I_E^{(2)}}.
\end{equation*}
Also, since $N(\omega)\sin\left(\omega\tau\right)\simeq\omega N(\omega)\tau\to2T_R\tau$, the last term of Eq.~\eqref{bsim} reads
\begin{equation*}
    8iT_R\Gamma_0^2\int^{+\infty}_{-\infty}d\tau\tau G_R\partial^2_{\tau}G_L=2iT_R\frac{\partial S_E^{(11)}(\omega)}{\partial \omega}\Bigg|_{\omega\to0}
\end{equation*}
where we have referred to Eq.~\eqref{11term} for the definition of $S_E^{(11)}(\omega)$. Overall, then, the excess heat current fluctuations in the DC limit has the following form,
\begin{equation}
    \boxed{S_E(0)=
 4T_R(\nu-1)\braket{I_E^{(2)}}+S_E^{(11)}(0)+2iT_R\frac{\partial S_E^{(11)}(\omega)}{\partial \omega}\Bigg|_{\omega\to0}}\label{shot noise main}
\end{equation}

Similarly to the heat current, we will evaluate $S_E(0)$ in two different limits and investigate the ratio between the heat current fluctuations and the heat current.
 
\subsection{Almost equal temperature limit}\label{app cft almost}
Similar to before, we consider a small temperature gradient between the two edges by defining $T_R=T$, $T_L=T+\Delta T$, and expanding the last two terms in Eq.~\eqref{shot noise main} up to the first order of $\Delta T/T$. 

\subsubsection{$S_E^{(11)}(0)$}
Let us first focus on $S_E^{(11)}(0)$. From Eq.~\eqref{derivative}, a simple calculation shows
\begin{equation}
    \partial_{\tau}^2G_R(\tau) = \nu(\pi T_R)^2\Biggl(\nu+\frac{1+\nu}{\sinh^2(\pi T_R(\tau-i\epsilon))}\Biggr)G_R.\label{seconddr}
\end{equation}
Therefore, referring to Eq.~\eqref{11term}, we obtain
\begin{eqnarray*}
 S_E^{(11)}(0)&=&-4\Gamma_0^2\int^{+\infty}_{-\infty}d\tau G_R(\tau)\partial_{\tau}^2G_{L}(\tau)=-4\Gamma_0^2\int^{+\infty}_{-\infty}d\tau (\partial^2_{\tau}G_R(\tau))G_{L}(\tau) \\
 &=&-
 \frac{4 \Gamma_0^2\nu}{\Lambda^{2\nu}}\int^{+\infty}_{-\infty}d\tau\frac{\nu(\pi T_R)^{\nu+2}(\pi T_L)^{\nu}}{[i\sinh(\pi T_R(\tau-i\epsilon))]^\nu[i\sinh(\pi T_L(\tau-i\epsilon))]^{\nu}} \\
 &+&\frac{4 \Gamma_0^2\nu}{\Lambda^{2\nu}}\int^{+\infty}_{-\infty}d\tau
 \frac{(1+\nu)(\pi T_R)^{\nu+2}(\pi T_L)^{\nu}}{[i\sinh(\pi T_R(\tau-i\epsilon))]^{\nu+2}[i\sinh(\pi T_L(\tau-i\epsilon))]^{\nu}}.
\end{eqnarray*}
Similarly to the previous section, we change variables as $\lambda = \frac{T_L}{T_R}= 1+\frac{\Delta T}{T}$, $z=\pi T_R(\tau-i\epsilon)+\frac{i \pi}{2}$. We then expand up to the first order of $\frac{\Delta T}{T}$, obtaining
\begin{equation}
    \begin{aligned}
        S_E^{(11)}(0)\simeq-
         \frac{4 \Gamma_0^2\nu^2(\pi T)^{2\nu+1}}{\Lambda^{2\nu}}\int^{+\infty}_{-\infty}dz\Bigl(1+\nu\frac{\Delta T}{T}\Bigr)\frac{1-\nu(z-\frac{i\pi}{2})\tanh \left( z \right)\frac{\Delta T}{T}}{[\cosh \left( z \right)]^{2\nu}} \\
         +
         \frac{4 \Gamma_0^2\nu(1+\nu)(\pi T)^{2\nu+1}}{\Lambda^{2\nu}}\int^{+\infty}_{-\infty}dz\Bigl(1+\nu\frac{\Delta T}{T}\Bigr)\frac{1-\nu(z-\frac{i\pi}{2})\tanh \left( z \right) \frac{\Delta T}{T}}{[\cosh \left( z \right)]^{2\nu+2}}.\label{exs11}
    \end{aligned}
\end{equation}
The contribution of order $\mathcal{O} \left( 1 \right)$ in Eq.~\eqref{exs11} is calculated to be
\begin{equation}
     \frac{4\Gamma_0^2}{\Lambda^{2\nu}}\frac{2^{2\nu-1}(\pi T)^{2\nu+1}}{2\nu+1}\frac{\Gamma^2(\nu+1)}{\Gamma(2\nu)},\label{orderzero}
\end{equation}
where again, $\Gamma (x)$ is the Euler gamma function.
Dropping the terms which are odd functions of $z$, the contribution of order $\mathcal{O} \left(\frac{\Delta T}{T} \right)$ reads
\begin{equation*}
    \frac{4 \Gamma_0^2 \nu^2 (\pi T)^{2\nu+1}} {\Lambda^{2\nu}} \int^{+\infty}_{-\infty}dz 
    \Biggl(-\frac{\nu}{[\cosh \left( z \right) ]^{2\nu}} + \frac{\nu z\sinh \left( z \right) }{[\cosh \left( z \right) ]^{2\nu+1}} + \frac{1+\nu}{[\cosh \left( z \right) ]^{2\nu+2}} -\frac{(1+\nu)z\sinh \left( z \right) }{[\cosh \left( z \right) ]^{2\nu+3}}\Biggr) \frac{\Delta T}{T}
\end{equation*}
Performing the integration, we arrive, after some algebra, at the final form
 \begin{equation}
 \boxed{
S_E^{(11)}(0)=  \frac{4\Gamma_0^2}{\Lambda^{2\nu}}\frac{2^{2\nu-1}(\pi T)^{2\nu+1}}{2\nu+1}\frac{\Gamma^2(\nu+1)}{\Gamma(2\nu)}+\frac{2\Gamma_0^22^{2\nu-1}(\pi T)^{2\nu+1}}{\Lambda^{2\nu}}\frac{\Gamma^2(\nu+1)}{\Gamma(2\nu)}\frac{\Delta T}{T}+\mathcal{O}\Bigl(\bigl(\frac{\Delta T}{T}\bigr)^2\Bigr).}\label{sum1}
      \end{equation}
\subsubsection{The derivative term}
Now we move on to the last term of the excess heat current fluctuations, as given in Eq.~\eqref{shot noise main}. The derivative of the noise $S_E^{(11)}(\omega)$ is described by
\begin{eqnarray}
i\frac{\partial S_E^{(11)}(\omega)}{\partial \omega}\Bigg|_{\omega\to0}=4\Gamma_0^2\int^{+\infty}_{-\infty}d\tau i\tau G_R(\tau)\partial^2_{\tau}G_{L}(\tau)
=4\braket{I_E^{(2)}}
+4\Gamma_0^2\int^{+\infty}_{-\infty}d\tau i\tau (\partial^2_{\tau}G_R(\tau))G_{L}(\tau),\label{derivative11}
 \end{eqnarray}
where the second equality is obtained we have performed partial integration several times. Referring to Eq.~\eqref{seconddr}, the last term of Eq.~\eqref{derivative11} is transformed into
\begin{eqnarray*}
   4\Gamma_0^2\int^{+\infty}_{-\infty}d\tau i\tau (\partial^2_{\tau}G_R(\tau))G_{L}(\tau) &=&
 \frac{4i \Gamma_0^2\nu}{\Lambda^{2\nu}}\int^{+\infty}_{-\infty}d\tau\frac{\tau\nu(\pi T_R)^{\nu+2}(\pi T_L)^{\nu}}{[i\sinh(\pi T_R(\tau-i\epsilon))]^\nu[i\sinh(\pi T_L(\tau-i\epsilon))]^{\nu}}\nonumber\\
 &-&\frac{4i \Gamma_0^2\nu}{\Lambda^{2\nu}}\int^{+\infty}_{-\infty}d\tau
 \frac{\tau(1+\nu)(\pi T_R)^{\nu+2}(\pi T_L)^{\nu}}{[i\sinh(\pi T_R(\tau-i\epsilon))]^{\nu+2}[i\sinh(\pi T_L(\tau-i\epsilon))]^{\nu}}\label{870}.
\end{eqnarray*} 

Implementing a similar expansion as in App.~\ref{app expansion}, and performing the integration, one finds
\begin{equation}
\boxed{
i\frac{\partial S_E^{(11)}(\omega)}{\partial \omega}\Bigg|_{\omega\to0}=  4\braket{I_E^{(2)}}-\frac{2\pi\Gamma_0^2}{\Lambda^{2\nu}}\frac{2^{2\nu-1}(\pi T)^{2\nu}}{2\nu+1}\frac{\Gamma^2(\nu+1)}{\Gamma(2\nu)}\Bigl(1+(\nu+1)\frac{\Delta T}{T}\Bigr)+\mathcal{O}\Bigl(\bigl(\frac{\Delta T}{T}\bigr)^2\Bigr)}.\label{sum2}
 \end{equation}
\subsubsection{ Excess heat current fluctuations and the heat Fano factor}
Combining Eqs.~\eqref{shot noise main}, \eqref{sum1}, and \eqref{sum2}, in the almost equal temperature limit, we finally arrive at
\begin{equation}
    \boxed{S_E(0)=4T(\nu+1)\braket{I_E^{(2)}}
-\frac{\Gamma_0^22^{2\nu-1}(\pi T)^{2\nu+1}}{\Lambda^{2\nu}}\frac{2}{2\nu+1}\frac{\Gamma^2(\nu+1)}{\Gamma(2\nu)}\frac{\Delta T}{T}+\mathcal{O}\Bigl(\bigl(\frac{\Delta T}{T}\bigr)^2\Bigr)}.\label{ex total}
\end{equation}
Recalling the form of the heat current from Eq.~\eqref{current1}, \begin{equation*}
    \braket{I_E^{(2)}}=\frac{\pi\Gamma_0^2(\pi T)^{2\nu}}{\Lambda^{2\nu}}\frac{2^{2\nu-1}}{2\nu+1}\frac{\Gamma^2(\nu+1)}{\Gamma(2\nu)}\frac{\Delta T}{T}+\mathcal{O}\Bigl(\bigl(\frac{\Delta T}{T}\bigr)^3\Bigr),
\end{equation*}
it immediately follows that
\begin{equation}
    \boxed{\frac{S_E(0)}{2\braket{I_E^{(2)}}}=(2\nu+1)T=(4h_{\nu}+1)T,}\label{ex total2}
\end{equation}
where we have utilized the scaling dimension of the Laughlin quasiparticle, $h_{\nu}=\nu/2$. We have thus completed the proof of Eq.~\eqref{eq:Close_Temp} in the main text in the case of Abelian FQH with filling fraction $\nu$. 

\subsection{Zero temperature limit}\label{boson zero 3}
We turn to the different limit of temperature where we set $T_R=0$, $T_L=T$.
The only surviving term in the excess heat current fluctuations Eq.~\eqref{shot noise main} becomes $S_E(0)=S_E^{(11)}(0)$, and the contributions from the cross terms are dropped. Recalling the correlator of the right-moving edge is given at zero temperature by $G_R(\tau) \rightarrow \Bigl[i\Lambda(\tau-i\epsilon)\Bigr]^{-\nu}$, and changing variables as usual by $z=\pi T(\tau-i\epsilon)+i\frac{\pi}{2}$, one finds
 \begin{equation}
 \boxed{
S_E(0)= \frac{4\Gamma_0^2}{\Lambda^{2\nu}}\int^{+\infty}_{-\infty}dz
\frac{\nu(\nu+1)(\pi T)^{2\nu+1}}{[\cosh(z)]^{\nu}[i(z-i\frac{\pi}{2})]^{\nu+2}}}.\label{b110}
      \end{equation}
Similarly to the heat current, it is challenging to analytically evaluate this form, except the case of integer values of $\nu$.
      
\subsubsection{An electron tunneling}
In the same way as the previous section, we can evaluate this form at $\nu=1$ which corresponds to the electron tunneling. The excess heat current fluctuations are now given by
\begin{equation*}
   S_E(0)= \frac{4\Gamma_0^2}{\Lambda^{2}}\int^{+\infty}_{-\infty}dz
\frac{2(\pi T)^{3}}{[\cosh(z)][i(z-i\frac{\pi}{2})]^{3}}.
\end{equation*}
By introducing the same contour as Fig.~\ref{fig2}, we apply the Cauchy theorem. By calculating the residue, we find
\begin{equation*}
   S_E(0)=- \frac{16 \pi \Gamma_0^2 T^3}{\Lambda^{2}}\sum_{n=1}^{\infty}\frac{(-1)^n}{n^3},
\end{equation*}
which combined with the identity $\sum_{n=1} ^{\infty} \frac{(-1)^n}{n^3}=-\frac{3}{4}\zeta(3)$ gives
\begin{equation}
  \boxed{ S_E(0)= \frac{12\Gamma_0^2}{\Lambda^{2}}\pi T^3\zeta(3)}.
\end{equation}
\subsubsection{Generic integer case and the heat Fano factor}
Similarly to App.~\ref{ap generic abelian}, one can evaluate the heat current fluctuations of Eq.~\eqref{b110} for generic integer values of $\nu$. 
For odd integer values of $\nu$, we have
\begin{equation}
S_E(0)=\frac{4\Gamma_0^2}{\Lambda^{2}}\nu(\nu+1)(\pi T)^{2\nu+1}\sum_{0\leq a\leq \frac{\nu-1}{2}}p^\prime_a\eta(\nu+1+2a)\label{b115}
\end{equation}
with $p^\prime_a=\frac{\nu+2a+1}{(\nu+1)\pi}p_a$, where $p_a$ is defined in Eq.~\eqref{b40}. For even integer values, we obtain
\begin{equation}
    \frac{4\Gamma_0^2}{\Lambda^{2}}\nu(\nu+1)(\pi T)^{2\nu+1}\sum_{1\leq b\leq \frac{\nu-1}{2}}q^\prime_b\zeta(\nu+2b)\label{b116}
\end{equation}
with $q_b^\prime=\frac{\nu+2b}{(\nu+1)\pi}q_b$ (see~\eqref{b41}). We can get the heat Fano factor by taking ratio between the noise and the current via $\mathcal{F}_E=\frac{S_E(0)}{2\braket{I_E^{(2)}}}$. From Eqs.~ \eqref{b40}\eqref{b41}\eqref{b115}\eqref{b116}, we find
\begin{equation}
    \label{eq:zero_temp_general_integer}
        \mathcal{F}_E=
    \begin{cases}
   T\frac{\sum_{0\leq a\leq \frac{\nu-1}{2}}(\nu+2a+1)p_a\eta(\nu+2+2a)}{\sum_{0\leq a\leq \frac{\nu-1}{2}}p_a\eta(\nu+1+2a)}\;\;(\nu\in2\mathbb{N}-1)\\
   T\frac{\sum_{1\leq b\leq \frac{\nu}{2}}(\nu+2b)q_b\zeta(\nu+2b+1)}{\sum_{1\leq b\leq \frac{\nu}{2}}q_b\zeta(\nu+2b)}\;\;(\nu\in2\mathbb{N})\end{cases}.
\end{equation}

\section{Generic CFTs}\label{App:generic_cft}
We can extend the previous discussions to a generic conformal field theory (CFT), especially to the case of a rational CFT (RCFT). These include the Ising and $SU(2)_k$ CFTs, corresponding to edge modes of the Moore-Read state~\cite{MOORE1991362} and $SU(2)_k$ topological order phases (spin liquid phases), respectively. Particularly, in the spin liquid phases, the excitations do not carry charge, therefore heat current and noise is an important probe for these excitations. We focus on the \textit{neutral} topological phases  whose edge mode is described by a RCFT with central charge $c$.
The essential difference from the previous bosonization formalism is that the heat current is given by the stress-energy tensor~\cite{CAPPELLI2002568}; as a consequence, to evaluate the noise, one needs to calculate correlator between the stress-energy tensor and primary fields. Compared to the bosonization case, this corresponds to the formula in Eq.~\eqref{b formula}. 
\par
\subsection{Heat current}
We consider the same geometry as Fig.~\ref{fig01} with the edge modes replaced by a RCFT with central charge $c$ and a tunneling of a neutral anyon through a QPC at $x=0$. This is given by the following Hamiltonian
\begin{eqnarray}
H&=&H_0+H_T\nonumber\\
  H_{0}&=&\frac{v}{2\pi}\int _{-\infty}^{+\infty}dx[\mathcal{T}+\overline{\mathcal{T}}]\nonumber\\
      H_{T}&=& \Gamma_0 O_R(0)O_L(0).\label{76}   
\end{eqnarray}
Here, $H_0$ describes kinetic terms consisting of the stress-energy tensor $\mathcal{T}(z)/\overline{\mathcal{T}}(\overline{z})$ in the right-/left-moving (\textit{i.e.,} holomorphic/anti-holomorphic) sector.
Also, $H_T$ represents the tunneling, where $O_{R/L}(z)$ is a primary field of the RCFT with scaling dimension $h_{O_R}=h_{O_L}\equiv h_{O}$ (see technical caveat~\cite{TechnicalCaveat} in the main text).  We interchangeably distinguish the two sectors by using the bar notation, namely, $O(z)$ and $\overline{O}(\overline{z})$ correspond to the right- and left-moving edge modes, respectively. 
\par
\subsubsection{Unperturbed heat current}
Before diving into the calculation of the perturbative expansion of the heat current, let us briefly review the heat current in the ground state~\cite{CAPPELLI2002568}.
The unperturbed heat current operator is given by 
\begin{equation}
  I_E^{(0)}(x,t)=\frac{v^2}{2\pi}\mathcal{T}(z).  \label{c2}
\end{equation}
To find the expectation value at finite temperature $T_R$, we implement a conformal mapping from the complex plane to a cylinder, which is accomplished by
 $z=\exp[2\pi iT_Rw/v]$. Under this transformation, the stress-energy tensor is transformed into
\begin{equation}
 \mathcal{T}(w)= \Bigl(\frac{dw}{dz}\Bigr)^{-2}\biggl[\mathcal{T}(z)-\frac{c}{12}S(w,z)  \biggr]\label{swrz}
\end{equation}
where $S(z,w)$ represents the Schwarzian derivative~\cite{DiFrancesco1997}:
\begin{equation}
    S(w,z)=\Bigl(\frac{\partial^3w}{\partial z^3}\Bigr)\Bigl(\frac{\partial w}{\partial z}\Bigr)^{-1}-\frac{3}{2}\Bigl(\frac{\partial^2w}{\partial z^2}\Bigr)^2\Bigl(\frac{\partial w}{\partial z}\Bigr)^{-2}.\label{swrz2}
\end{equation}
The Schwarzian derivative becomes $ S(w,z)=\frac{1}{2}e^{-4\pi iT_Rw/v}$ in the present case. Taking the expectation value on the both sides of \eqref{swrz} and using $\braket{\mathcal{T}(z)}=0$ leads to
\begin{equation*}
\braket{\mathcal{T}(w)}=\frac{\pi^2cT_R^2}{6v^2},
\end{equation*}
hence,
\begin{equation*}
    \braket{ I_E^{(0)}(x,t)}=\frac{\pi cT_R^2}{12}.
\end{equation*}
Remarkably, this quantity was observed for both integer~\cite{jezouin2013quantum,banerjee_observed_2017} and half-integer~\cite{banerjee_observation_2018} central charges through heat conductance experiments.

\subsubsection{Perturbative expansion}
As we did previously, to get the perturbative expansion of the heat current, we need to find a commutation relation between the  stress energy tensor $\mathcal{T}(z)$ associated with the the energy operator, and the primary field $O(z^{\prime})$ in the holomorphic sector. To this end, we resort to the operator product expansion~\cite{DiFrancesco1997} between the stress energy tensor $\mathcal{T}(z)$ and a primary field $O(z^{\prime})$:
\begin{eqnarray*}
    \mathcal{T}(z)O(z^{\prime})\sim\frac{h_{O}O(z^{\prime})}{(z-z^{\prime})^2}+\frac{\partial_{z^{\prime}} O(z^{\prime})}{z-z^{\prime}}+\cdots\nonumber\\
   O(z)\mathcal{T}(z^{\prime})\sim\frac{h_{O}O(z)}{(z-z^{\prime})^2}-\frac{\partial_z O(z)}{z-z^{\prime}}+\cdots.  \end{eqnarray*}
Using this relation, and changing variables from complex to real time by replacing $z-z^{\prime}\to\delta+iv(t_1-t_2)$, with $\delta$ being an infinitesimal number, we have
\begin{eqnarray*}
    [\mathcal{T}(t_1),O(t_2)]&=&\frac{h_{O}O(t_2)}{(iv)^2(t_1-t_2-i\delta)^2}+\frac{\partial_{t_2}O(t_2)}{(iv)^2(t_1-t_2-i\delta)}-\frac{h_{O}O(t_2)}{(iv)^2(t_2-t_1-i\delta)^2}+\frac{\partial_{t_2}O(t_2)}{(iv)^2(t_2-t_1-i\delta)}\nonumber\\
    &=&\frac{1}{(iv)^2}\Biggl[\frac{1}{t_1-t_2-i\delta}-\frac{1}{t_1-t_2+i\delta}\Biggr]\partial_{t_2} O(t_2)-\frac{h}{(iv)^2}\partial_{t_1}\Biggl[\frac{1}{t_1-t_2-i\delta}-\frac{1}{t_1-t_2+i\delta}\Biggr]O(t_2)\nonumber\\
    &=&-\frac{2\pi i}{v^2}\delta(t_1-t_2)\partial_{t_2} O(t_2)+\frac{2\pi i}{v^2}h_O\partial_{t_1}(\delta(t_1-t_2))O(t_2).\label{comm}
\end{eqnarray*}
In the last equality, we have used the relation
\begin{equation*}
   \Bigl[\frac{1}{x+i\delta}-\frac{1}{x-i\delta}\Bigr]\Bigg|_{\delta\to0^+}=-2\pi i\delta(x).
\end{equation*}

We introduce the heat current operator at the drain and time $t$ by
$I_E(d,t)=\frac{v^2}{2\pi}\mathcal{T}(\tilde{t})$ with $\tilde{t}=t-d/v$,
similarly to Eqs.~\eqref{IE}-\eqref{2nd}. The first order of the heat current is given by
\begin{equation}
    \begin{aligned}
      I_E^{(1)}(t) & =i\int^t_{-\infty}dt^\prime[H_T(t^{\prime}),I_E(\tilde{t})]=-i\Gamma_0\int^t_{-\infty}dt^{\prime}[I_E(\tilde{t}),O(t^{\prime})]\overline{O}(t^{\prime}) \\
      &=\Gamma_0\int^t_{-\infty}dt^\prime\Bigl(h_O\partial_t\delta(\tilde{t}-t^{\prime})O(t^{\prime})\overline{O}(t^{\prime})-\delta(\tilde{t}-t^{\prime})(\partial_{t^\prime}O(t^{\prime}))\overline{O}(t^{\prime})\Bigr) \\
       &=\Gamma_0\Bigl(h_O\partial_{t}(O(\tilde{t})\overline{O}(\tilde{t}))-(\partial_tO(\tilde{t}))\overline{O}(\tilde{t})\Bigr).\label{yy}
    \end{aligned}
\end{equation}
The first term of \eqref{yy} becomes zero when taking an expectation value; also, in the second order perturbations, the first term contributes an integral over a total derivative of a correlator,~$\sim\int d\tau\partial_{\tau}(G_R(\tau)G_L(\tau))$. Thus, it can be omitted entirely.
What remains is
\begin{equation}
    I_E^{(1)}(t)=-\Gamma_0(\partial_tO(\tilde{t}))\overline{O}(\tilde{t})=\Gamma_0O(\tilde{t})\partial_t\overline{O}(\tilde{t}).
\end{equation}
Taking the expectation value, one finds $\braket{I_E^{(1)}(t)}=0$, \textit{i.e.,} there is no first order correction.\par 
The second order correction of the heat current is given by
\begin{equation}
    \begin{aligned}
     I_E^{(2)}(t) & =i\Gamma_0^2\int^{\tilde{t}}_{-\infty}dt^{\prime\prime}[O_R(t^{\prime\prime})O_L(t^{\prime\prime}),O_R(\tilde{t})\partial_tO_L(\tilde{t})] \\
    &=i\Gamma_0^2\int^{\tilde{t}}_{-\infty}dt^{\prime\prime} \Bigl(O_R(t^{\prime\prime})O_R(\tilde{t})O_L(t^{\prime\prime})\partial_tO_L(\tilde{t})-O_R(\tilde{t})O_R(t^{\prime\prime})\partial_tO_L(\tilde{t})O_L(t^{\prime\prime}) \Bigr).\label{cftg2}
    \end{aligned}
\end{equation}
Defining 
\begin{equation}
 G_{R/L}(t_{12})=\braket{O_{R/L}(t_1)O_{R/L}(t_2)}=\Biggl[\frac{(\pi T_{R/L}/(i\Lambda))}{\sinh({\pi T_{R/L}(t_{12}-i\epsilon))}}\Biggr]^{2h_O}   \label{cft correlator}
\end{equation}
and $\tau=t^{\prime\prime}-\tilde{t}$, after some algebra, we get
\begin{equation}
   \boxed{\braket{ I_E^{(2)}(t)}= -i\Gamma_0^2\int^{+\infty}_{-\infty}d\tau G_R(\tau)\partial_{\tau}G_L(\tau)}.\label{cft2nd}
\end{equation}
As we mentioned in the footnote~\cite{TechnicalCaveat}, the amplitude of \eqref{cft2nd} is halved compared with the charged anyon case~\eqref{2nd2} (Abelian bosonization case~\ref{App:luttinger_liquid}) simply because we add the conjugate term in the tunneling Hamiltonian in the latter case~\eqref{1}. Such a discrepancy of the amplitude being halved is also seen in the heat current fluctuation, thus the heat Fano factor, defined by ratio between these two, is not influenced by this fact.   
\subsection{Heat current fluctuations}
We will study the heat current fluctuations, and their ratio with the heat current, in the DC limit. 
We begin with the heat current fluctuations, defined by
\begin{equation}
    \tilde{S}_E(\omega)=\int_{-\infty}^{+\infty}dt_{12}e^{i\omega t_{12}}\braket{\{\Delta I_E(t_1),I_E(t_2)\}},
\end{equation}
with $\Delta I_E(t)=I_E(t)-\braket{I_E(t)}$.
Expanding the heat current in orders of $\Gamma_0$, this is decomposed into
\begin{equation}
     \tilde{S}_E(\omega)=S_E^{(00)}(\omega)+S_E^{(11)}(\omega)+S_E^{(02)}(\omega)+S_E^{(20)}(\omega)+\mathcal{O}(\Gamma_0^3).\label{decompose}
\end{equation}
Here, we have defined
\begin{equation}
     S_E^{(ij)}(\omega)=\int_{-\infty}^{+\infty}dt_{12}e^{i\omega t_{12}}\braket{\{\Delta I^{(i)}_E(t_1),I^{(j)}_E(t_2)\}} \;(i,j=0,1,2).
\end{equation}
Notice that the term $S_E^{(01)}(\omega)$ and $S_E^{(10)}(\omega)$ are absent due to the charge neutrality condition. 
\subsubsection{Johnson-Nyquist noise of heat current}
Let us make a few comments on the first term
in Eq.~\eqref{decompose}, corresponding to the so-called Johnson-Nyquist noise of the heat current~\cite{Johnson,Nyquist}. Taking the correlation function of the stress-energy tensor in the complex plane,
\begin{equation}
    \braket{\mathcal{T}(z)\mathcal{T}(z^\prime)}=\frac{c}{2}\frac{1}{(z-z^\prime)^4},
\end{equation}
and implementing the conformal transformation~\eqref{swrz} allows us to obtain the fluctuations of the unperturbed heat current:
\begin{equation}
\braket{\{\Delta I^{(0)}_E(t_1),I^{(0)}_E(t_2)\}}=\frac{c}{4\pi^2}\biggl[\frac{\pi T_R}{\sinh(\pi T_R(t_{12}-i\delta))}\biggr]^4+\biggl(\frac{\pi cT_R}{12}\biggr)^2.
\end{equation}
After Fourier transformation, we get 
\begin{equation}
    S_E^{(00)}(\omega)=\frac{c\pi T_R^2}{6}\omega N(\omega)\biggl[1+\biggl(\frac{\omega}{2\pi T_R}\biggr)^2\biggr]=\kappa T_R\omega N(\omega)\biggl[1+\biggl(\frac{\omega}{2\pi T_R}\biggr)^2\biggr]\label{heat}
\end{equation}
with $\kappa$ being the heat conductance $\kappa=\frac{\partial \braket{I^{(0)}_E(t)}}{\partial T_R}=\frac{\pi^2c}{6}T_R$. In the limit of $\omega\to0$, Eq.~\eqref{heat} becomes 
\begin{equation*}
    S_E^{(00)}(0)=2\kappa T_R^2,
\end{equation*}
which is in close analogy to the fact that charge Johnson-Nyquist noise $S_c$ and the charge conductance $G$ is related via $S_c=2GT$ in the thermal noise limit. 
\subsubsection{Excess noise}

We focus on the last three terms of Eq.~\eqref{decompose}, representing the heat current fluctuations induced by the neutral anyon tunneling at the QPC. Using similar manipulations as in App.~\ref{app:auto_boson}, the second term of Eq.~\eqref{decompose} is given by
\begin{equation}
    \boxed{ S_E^{(11)}(\omega)=-2i\Gamma_0^2\int^{+\infty}_{-\infty}d\tau \cos \left( \omega\tau \right) G_R(\tau)\partial_{\tau}^2G_L(\tau)}.
\end{equation}
\subsubsection{conformal Ward identity}
Let us now analyze the cross terms. Initially we concentrate on $S_E^{(02)}(\omega)$. From Eqs.~\eqref{c2},\eqref{cftg2}, we explicitly write the Fourier transformation of  $\braket{\Delta I^{(0)}_E(t_1)I^{(2)}_E(t_2)}$ as
\begin{eqnarray}
    \int_{-\infty}^{+\infty}dt_{12}e^{i\omega t_{12}}\braket{\Delta I^{(0)}_E(t_1)I^{(2)}_E(t_2)}&=&-i\Gamma_0^2\int_{-\infty}^{+\infty}dt_{12}e^{i\omega t_{12}}\int_{-\infty}^{\tilde{t}_2}dt^{\prime\prime}\braket{\Delta I^{(0)}_E(\tilde{t}_1)O_R(t^{\prime\prime})O_R(\tilde{t}_2)}\partial_{t^{\prime\prime}}G_L(t^{\prime\prime}-\tilde{t}_2)\nonumber\\
    &+&i\Gamma_0^2\int_{-\infty}^{+\infty}dt_{12}e^{i\omega t_{12}}\int_{-\infty}^{\tilde{t}_2}dt^{\prime\prime}\braket{\Delta I^{(0)}_E(\tilde{t}_1)O_R(\tilde{t}_2)O_R(t^{\prime\prime})}\partial_{t^{\prime\prime}}G_L(\tilde{t}_2-t^{\prime\prime})\label{cft02}.
\end{eqnarray}

To evaluate the correlator involving the holomorphic sector, 
we make use of the conformal Ward identity.
This identity in complex plane is described by~\cite{DiFrancesco1997}
\begin{equation}
    \braket{\mathcal{T}(z)O_1(z_1)\cdots O_n(z_n)}=\sum_{i=1}^{n}\Biggl[\frac{h_i}{(z-z_i)^2}+\frac{1}{z-z_i}\partial_{z_i}\Biggr]  \braket{O_1(z_1)\cdots O_n(z_n)}\label{ward}
\end{equation}
where $O_i(z_i)$ is a primary field whose scaling dimension is given by $h_i$. We need to get a finite temperature correlator.
In practice, this can be accomplished by a conformal mapping from the complex plane to a cylinder via $z=\exp[2\pi iT_Rw/v]$. By this transformation, the stress-energy tensor, each primary field, and the derivative operator change as
\begin{equation*}
   \mathcal{T}(z)\to \Bigl(\frac{dw}{dz}\Bigr)^{2}\mathcal{T}(w)+\frac{c}{12}S(w,z)\;,  O_i(z_i)\to \Bigl(\frac{dw_i}{dz_i}\Bigr)^{h_i}O_i(w_i),\;\partial_{z_i}\to\frac{v}{2\pi iT_R}e^{-\frac{2\pi iT_R}{v}w_i}\partial_{w_i},
\end{equation*}
where $S(w,z)$ is the Schwarzian derivative  presented in Eq.~\eqref{swrz2}.
The Schwarzian derivative is $ S(w,z)=\frac{1}{2}e^{-4\pi iT_Rw/v}$ in the present case. After the conformal mapping, Eq.~\eqref{ward} becomes
\begin{eqnarray}
   \braket{\mathcal{T}(w)O_1(w_1)\cdots O_n(w_n)}&=&\sum_{i=1}^{n}\Biggl[\frac{h_i(2\pi i T_R)^2/v^2}{(1-e^{-\frac{2\pi iT_R}{v}(w-w_i)})^2}+\frac{e^{\frac{2\pi iT_R}{v}(w-w_i)}(2\pi iT_R)/v}{1-e^{-\frac{2\pi iT_R}{v}w_i}}\partial_{w_i}\Biggr]  \braket{O_1(w_1)\cdots O_n(w_n)}\nonumber\\
   &-&\frac{c}{24}e^{-4\pi iT_Rw/v}\braket{O_1(w_1)\cdots O_n(w_n)}
  \label{ward2}    
\end{eqnarray}

We investigate each term in the r.h.s of Eq.~\eqref{ward2}.
Indeed, some terms can be dropped. Since we will Fourier transform the functions in front of the correlators, we keep only the functions which have singularities (non-singular functions become zero) when $w$ is close to $w_i$. Also, due to the symmetry of the correlator $\braket{\mathcal{T}(w)O_1(w_1)\cdots O_n(w_n)}=\braket{O_1(w_1)\cdots O_n(w_n)\mathcal{T}(w)}$, the function has to be even under the exchange $w\leftrightarrow w_i$. Let us look at the first term of Eq.~\eqref{ward2}.
Setting $\theta_i=\frac{\pi iT_R}{v}(w-w_i)$, we have 
\begin{align*}
    \frac{1}{(1-e^{-\frac{2\pi iT_R}{v}(w-w_i)})^2}& = -\frac{1}{4}\cot^2\theta_i-\frac{i}{2}\cot\theta_i+\frac{1}{4}, \\
    \frac{e^{\frac{2\pi iT_R}{v}(w-w_i)}}{1-e^{-\frac{2\pi iT_R}{v}w_i}} & =-\frac{i}{2}\cot\theta_i+(\text{non-singular terms}).
\end{align*}
Due to the argument of the singularity and symmetry mentioned above, we can keep only the first term in each of these two lines (unlike the first line, in the second line the term $\cot \theta_i$ is combined with the derivative $\partial_{w_i}$, and hence it is even under the exchange of $w\leftrightarrow w_i$). A similar line of argument allows us to omit the last term of Eq.~\eqref{ward2}. 

After dropping terms, we further rotate
Eq.~\eqref{ward2} to real time by changing variables as $w-w_i\to\delta+iv(t-t_i)$, with $\delta$ being an infinitesimally small regulator. This yields
\begin{equation*}
    \braket{\mathcal{T}(t)O_1(t_1)\cdots O_n(t_n)}=\frac{2\pi}{v^2}\sum_{i=1}^{n}\Bigl[h_iP(t-t_i)+iK(t-t_i)\partial_{t_i}\Bigr]  \braket{O_1(t_1)\cdots O_n(t_n)},
\end{equation*}
where
\begin{subequations}
    \begin{align}
        P(t)&=-\frac{\pi  T_R^2}{2}\coth^2(\pi T_R(t-i\delta))\label{p}\\
    K(t)&=-\frac{T_R}{2}\coth(\pi T_R(t-i\delta)).\label{k}
    \end{align}
\end{subequations}
Identifying $\Delta I_E^{(0)}(t)=\frac{v^2}{2\pi}\mathcal{T}(t)$, we  arrive at
\begin{equation}
    \braket{\Delta I_E^{(0)}(t)O_1(t_1)\cdots O_n(t_n)}=\sum_{i=1}^{n}\Bigl[h_iP(t-t_i)+iK(t-t_i)\partial_{t_i}\Bigr]  \braket{O_1(t_1)\cdots O_n(t_n)}\label{ward4}.
\end{equation}

Coming back to evaluation of Eq.~\eqref{cft02}, we utilize the conformal Ward identity Eq.~\eqref{ward4} to find
\begin{equation}
    \begin{aligned}
   \eqref{ward4} = &-i\Gamma_0^2\int_{-\infty}^{0}d\tau\int_{-\infty}^{+\infty}dt_{12}e^{i\omega t_{12}}[h_OP(t_{12})-K(t_{12})\partial_{\tau}+h_OP(t_{12}-\tau)+K(t_{12}-\tau)\partial_{\tau}]G_R\partial_{\tau}G_L\\
    &-i\Gamma_0^2\int_{0}^{+\infty}d\tau\int_{-\infty}^{+\infty}dt_{12}e^{i\omega t_{12}}[h_OP(t_{12})+K(t_{12})\partial_{\tau}+h_OP(t_{12}+\tau)-K(t_{12}+\tau)\partial_{\tau}]G_R\partial_{\tau}G_L\label{cft02pt2}.
    \end{aligned}
\end{equation}
By Fourier transforming Eqs.~\eqref{p} and \eqref{k}, 
\begin{eqnarray*}
    P(\omega)=\frac{\omega}{2} N(\omega),\;K(\omega)=-\frac{i}{2}N(\omega)
\end{eqnarray*}
with $N(\omega)$ being
\begin{equation*}
   N(\omega)=
   \coth\Bigl(\frac{\omega}{2T_R}\Bigr)+1,
\end{equation*}
it follows that 
\begin{eqnarray}
    \eqref{cft02pt2}&=&-i\Gamma_0^2\int_{-\infty}^{0}d\tau[P(\omega)(1+e^{i\omega\tau})h_OG_R\partial_{\tau}G_L-K(\omega)(1-e^{i\omega\tau})\partial_{\tau}G_R\partial_{\tau}G_L]\nonumber\\
    &-&i\Gamma_0^2\int_{0}^{+\infty}d\tau[P(\omega)(1+e^{-i\omega\tau})h_OG_R\partial_{\tau}G_L+K(\omega)(1-e^{-i\omega\tau})\partial_{\tau}G_R\partial_{\tau}G_L]\label{cft02pt3}.
\end{eqnarray}

A similar line of arguments shows that the Fourier transformation of $\braket{I^{(2)}_E(t_2)\Delta I^{(0)}_E(t_1)}$ has the same integral expression as Eq.~\eqref{cft02pt3}. Hence, 
\begin{eqnarray}
    S_E^{(02)}(\omega)=&-&2i\Gamma_0^2\int_{-\infty}^{0}d\tau[P(\omega)(1+e^{i\omega\tau})h_OG_R\partial_{\tau}G_L-K(\omega)(1-e^{i\omega\tau})\partial_{\tau}G_R\partial_{\tau}G_L]\nonumber\\
    &-&2i\Gamma_0^2\int_{0}^{+\infty}d\tau[P(\omega)(1+e^{-i\omega\tau})h_OG_R\partial_{\tau}G_L+K(\omega)(1-e^{-i\omega\tau})\partial_{\tau}G_R\partial_{\tau}G_L]\label{cft02pt4}.
\end{eqnarray}
\par
The other cross term $S_E^{(20)}(\omega)$ can be evaluated in the similar fashion. Making use of the Kubo-Martin-Schwinger relation described in Eq.~\eqref{kms} 
(or direct calculation) yields
\begin{eqnarray}
    S_E^{(02)}(\omega)+S_E^{(20)}(\omega)&=&-2i\Gamma_0^2\int_{-\infty}^{+\infty}d\tau [(1+\cos\omega\tau)h_O\omega N(\omega)G_R\partial_{\tau}G_L+N(\omega)\sin\omega\tau \partial_{\tau}G_R\partial_{\tau}G_L]\nonumber\\
  &=&   -2i\Gamma_0^2\int_{-\infty}^{+\infty}d\tau [(1+\cos\omega\tau)h_O\omega N(\omega)G_R\partial_{\tau}G_L-\omega N(\omega)\cos\omega\tau G_R\partial_{\tau}G_L-N(\omega)\sin\omega\tau G_R\partial^2_{\tau}G_L]\nonumber.
\end{eqnarray}
In the last equality, we have implemented partial integration. 
\subsection{Excess heat current fluctuation}
In the DC limit, $\omega\to 0$, referring to Eqs.~\eqref{cft2nd},\eqref{decompose}, we find that the excess heat current fluctuations $S_E(0)=\tilde{S}_E(0)-S_E^{(00)}(0)$ has the form
\begin{equation}
    \boxed{S_E(0)=
 4T_R(2h_O-1)\braket{I_E^{(2)}}+S_E^{(11)}(0)+2iT_R\frac{\partial S_E^{(11)}(\omega)}{\partial \omega}\Bigg|_{\omega\to0}}. \label{shot noise main cft}
\end{equation}
\subsection{Almost equal temperature limit}
Similarly to the case of bosonization, 
in the almost equal temperature limit, 
we can expand Eq.~\eqref{shot noise main cft} in orders of $\Delta T/T$ in a manner similar to the previous section (see App.~\ref{app cft almost}).
After the expansion, we arrive at
\begin{equation}
    \boxed{\frac{S_E(0)}{2\braket{I_E^{(2)}}}=(4h_O+1)T}, 
    \end{equation}
which is nothing but Eq.~\eqref{eq:Close_Temp} in the main text when we retrieve the Boltzmann constant $k_B$.
   
   \subsection{Zero temperature limit}
   \label{app:CFT_zero_temp}
   We can calculate the heat Fano factor in the zero temperature limit by setting 
   $T_R=T$ and $T_L=0$. The argument in this limit closely parallels the one presented in the previous sections, especially sections Apps.~\ref{boson zero 2} and \ref{boson zero 3}. The Fano factor has the integral expression reading
   \begin{equation}
   \mathcal{F}_E=(2h_O+1)(\pi T)
   \frac{J \left(2 h_O,2 \right)}{J \left(2 h_O,1 \right)},
    \label{integral_ex_compact app}
\end{equation}
where we define the integral
\begin{equation*}
    J \left(a,b\right) =  \int^{+\infty}_{-\infty}dz
    \left[ \cosh(z) \right]^{-a} 
    \left[i \left( z-i\frac{\pi}{2} \right) \right]^{- a - b}.
\end{equation*}

The form Eq.~\eqref{integral_ex_compact app} can be explicitly calculated when the scaling dimension $h_O$ is half-integer or integer. For the half-integer case, 
    \begin{equation}
    \label{eq:CFT_zero_temp_half_ints}
        \mathcal{F}_E=T\frac{\sum_{0\leq a\leq h_O-1/2}(2h_O+2a+1)p_a\eta(2h_O+2+2a)}{\sum_{0\leq a\leq h_O-1/2}p_a\eta(2h_O+1+2a)}\;\;(h_O\in\mathbb{N}-1/2),
 \end{equation}
where the coefficient $p_a$ is defined by
\begin{subequations}
     \begin{align}
        p_{h_O-1/2}&=\frac{2}{\pi^{h_O-1}}\binom{4h_O-1}{2h_O} \\
        p_a&=\frac{2}{\pi^{2h_O+2a}}\binom{2h_O+2a}{2a}\sum_{\substack{k\geq 1\\}} 
        \sum_{\substack{\{m_{l}\}\geq1\\1\leq l\leq k\\
        2\sum_lm_l=2h_O+1-2a
        }}
        \binom{2h_O+k-1}{k}\frac{(-1)^{k+a+h_O-1/2}}{\prod_{l}(2m_{l}+1)!}
        \;(0\leq a \leq h_O-\frac{3}{2})         
     \end{align}
\end{subequations}
and $\eta(s)$ is again the Dirichlet eta function. For integer case of $h_O$ we have
\begin{equation}
    \label{eq:CFT_zero_temp_ints}
    \mathcal{F}_E= T\frac{\sum_{1\leq b\leq h_O}(2h_O+2b)q_b\zeta(2h_O+2b+1)}{\sum_{1\leq b\leq h_O}q_b\zeta(2h_O+2b)}\;\;(h_O\in\mathbb{N}),
\end{equation}
where the coefficient $q_b$ is introduced by
\begin{subequations}
     \begin{align}       q_{h_O}&=\frac{2}{\pi^{h_O-1}}\binom{4h_O-1}{2h_O}\\
        q_b&=\frac{2}{\pi^{2h_O+2b-1}}\binom{2h_O+2b-1}{2b}\sum_{\substack{k\geq 1\\}} 
        \sum_{\substack{\{m_{l}\}\geq 1\\1\leq l\leq k\\
        2\sum_lm_l=2h_O-2b
        }}
        \binom{2h_O+k-1}{k}\frac{(-1)^{k+b+h_O}}{\prod_{l}(2m_{l}+1)!}\;\;(1\leq b \leq h_O-1)
     \end{align}
\end{subequations}
We give several examples of the heat Fano factor:
\begin{eqnarray*}
\mathcal{F}_E&=&2T\frac{\eta(3)}{\eta(2)}=3T\frac{\zeta(3)}{\zeta(2)}\;\;\;(h_O=1/2)\\
\mathcal{F}_E&=&4T\frac{\zeta(5)}{\zeta(4)}\;\;\;(h_O=1)\\
\mathcal{F}_E&=&4T\frac{30\eta(7)+\pi^2\eta(5)}{20\eta(6)+\pi^2\eta(4)}=15T\frac{63\zeta(7)+32\pi^2\zeta(5)}{155\zeta(6)+7\pi^2\zeta(4)}\;\;\;(h_O=3/2)\\
\mathcal{F}_E&=&5T\frac{112\zeta(9)+6\pi^2\zeta(7)}{70\zeta(8)+5\pi^2\zeta(6)}\;\;\;(h_O=2).
\end{eqnarray*}

\end{widetext}
\end{document}